%% file: main.tex
\documentclass[]{relaxed_system_lab}

\usepackage[utf8]{inputenc}
\usepackage[T1]{fontenc}
\usepackage{charter}
\usepackage{varwidth}
\usepackage{wrapfig}
\usepackage{fontawesome5}

\usepackage{multirow}
\usepackage{booktabs}       
\usepackage[table]{xcolor}  
\usepackage{array}
\usepackage{tabularx}

\newcolumntype{Y}{>{\centering\arraybackslash}X}

\usepackage{amsmath, amssymb, amsfonts, bm} 
\usepackage{graphicx}          

\usepackage{xspace}            
\usepackage{enumitem}          

\usepackage{makecell}          

\usepackage{algorithm}
\usepackage{algpseudocode}

\usepackage{float}
\usepackage{soul}

\input{macro}

\definecolor{lightgreen}{RGB}{144,238,144}
\definecolor{darkcyan}{HTML}{008B8B}

\newcommand{\sys}{\textsc{AReaL-DTE}\xspace}

\definecolor{gsmDark}{HTML}{315B73}
\definecolor{humanDark}{HTML}{39706D}
\definecolor{mathDark}{HTML}{665A7A}
\definecolor{logiqaDark}{HTML}{8A653B}

\definecolor{gsmLight}{HTML}{E8F1F6}
\definecolor{humanLight}{HTML}{E7F2F0}
\definecolor{mathLight}{HTML}{F0EDF5}
\definecolor{logiqaLight}{HTML}{F7F0E7}

\definecolor{oursRow}{HTML}{EAF3FF}
\definecolor{alternateRow}{HTML}{F7F8FA}
\definecolor{oursAccent}{HTML}{2563A6}

\definecolor{headerGray}{HTML}{F1F2F3}
\definecolor{subheaderGray}{HTML}{F8F8F8}

\title{\sys: Sparse Policy-Weight Transfer for Online Agentic Reinforcement Learning}

\author{Yingqi Peng$^{1,2}$ , Jiawei Zhang$^2$, Wenhao Zhou$^{1,2}$, Ruida Xu$^2$, Ran Yan$^{2,3}$, Wei Dong$^1$, Yi Gao$^{1,*}$, Zhiqiang Ding$^{2,*}$, \\
Tongkai Yang$^2$, Binhang Yuan$^{2,3,*}$}

\affiliation{$^1$Zhejiang University, $^2$Ant Group, $^3$HKUST}
\vspace{-1.0em}

\abstract{
Online agentic reinforcement learning implemented with micro-services separates policy training from rollout generation, improving scalability and modularity while potentially making frequent policy-weight synchronization a critical systems overhead. Shared storage naturally connects these services across clusters, but vanilla dense policy weight synchronization could incur model-scale construction, transfer, and application costs. Sparse synchronization reduces transferred data, yet checkpoint-oriented approaches can still retain a previous model and materialize complete intermediates to bridge heterogeneous training and inference layouts.
We present \sys, a snapshot-free \underline{D}elta \underline{T}ransfer \underline{E}ngine that translates inference-visible weight sparsity into end-to-end system efficiency. Across our evaluated workloads, fewer than $2\%$ of \texttt{BF16} weight elements change between consecutive policy versions. \sys reconstructs overwritten weights on demand by inverting AdamW updates, streams reconstructed and current parameters through converter-aligned \texttt{BF16} change detection, and remaps changed elements directly into receiver-local coordinates. \sys supports manifest-committed sparse transfer through shared storage across clusters and a deadlock-safe two-round protocol within a cluster, followed by direct application to inference shards.
We evaluate \sys on Qwen3-8B and Qwen3-30B-A3B across four online RL workloads. \sys achieves speedups of up to $19.9\times$ over \textsc{ByteCheckpoint} and $3.2\times$ over \textsc{PULSE} across clusters, and up to $7.6\times$ and $7.4\times$, respectively, within a cluster. In the same-cluster Qwen3-30B-A3B experiments, it reduces peak GPU memory by approximately $41\%$ and peak CPU memory by at least $87\%$.

\vspace{-2.0em}
}

\begin{document}
\maketitle

\begin{center}
\vspace{-1.5em}
\href{https://github.com/areal-project/AReaL-DTE}{%
  \faGithub\,\texttt{\:Code: https://github.com/areal-project/AReaL-DTE}}
\vspace{0.5em}
\end{center}

\section{Introduction}
\label{sec:intro}

Online reinforcement learning (RL), in which a policy model repeatedly generates trajectories, receives rewards, and updates its policy, has become an important mechanism for advancing LLM capabilities for production-level deployment~\cite{cursor2026realtimeRL}. Supporting this learning paradigm requires a continuous systems loop connecting agent interaction, rollout generation, reward computation, and policy training. As these workloads grow, emerging systems increasingly deploy the stages as independently scalable components. For example, \textsc{AReaL-2.0} exposes agent, inference, training, and weight-synchronization services~\cite{yan2026next}; \textsc{ProRL-Agent} provides rollout as an independent service~\cite{zhang2026prorl}; and \textsc{RollArt}~\cite{gao2026rollart} and \textsc{FlexMARL}~\cite{jiang2026rollout} disaggregate rollout, environment execution, reward computation, and training across heterogeneous resources. This efficient \textit{micro-service-oriented infrastructure} improves scalability and modularity, but it also raises a critical problem: \textit{whenever training workers produce a new version of policy model, how can the weight-synchronization service efficiently transfer its inference-usable weights to rollout workers?}

Consistent shared storage provides a natural rendezvous for these decoupled micro-services. During agentic RL training, the training service writes an updated weight version to shared storage, and rollout services retrieve it when synchronizing their local policy. Such a storage-mediated path allows training and rollout workers to operate with independent lifecycles, communication groups, and cluster placements. Without careful optimization, however, each policy update must be constructed, staged, written, transferred, read, and installed. Representing every update as a complete checkpoint makes both transfer volume and local processing overhead grow linearly with model size, potentially dominating the interval between policy updates.

On the other hand, sparse weight synchronization offers a promising alternative. RL optimization typically changes only a small fraction of the floating-point values in the policy model visible to rollout inference worker, i.e., fewer than $2\%$ change in our evaluated workloads. Systems such as \textsc{PULSE}~\cite{miahi2026understanding} and SparrowRL~\cite{ruan2026rl} exploit this property by transferring changed indices and their new values rather than the complete policy. They also demonstrate how sparse updates can be exchanged through shared storage or represented as versioned delta checkpoints. Compared with dense checkpoint mechanisms such as ByteCheckpoint~\cite{wan2025bytecheckpoint}, these approaches substantially reduce the number of bytes transferred between training and rollout inference.

Checkpoint-oriented approaches such as \textsc{PULSE}~\cite{miahi2026understanding} and \textsc{SparrowRL}~\cite{ruan2026rl} successfully exploit update sparsity to reduce transfer volume, but a sparse payload does not eliminate the system overhead of constructing, routing, and applying it due to the following challenges. \underline{\textbf{First}}, because AdamW updates policy weights in place, conventional change detection must retain a model-scale copy of the previous weights for comparison. \underline{\textbf{Second}}, training and rollout inference engines may fuse, permute, rename, and shard the same parameters differently. A changed index in a training shard therefore cannot be directly applied to an inference shard; bridging these layouts through checkpoints can introduce full-state gathering, model-scale conversion, CPU--GPU staging, and resharding. \underline{\textbf{Third}}, after sparse changes are routed to their destination shards, the payload size for each sender-receiver pair is data-dependent and may be zero, complicating buffer allocation and matched communication. Efficient sparse synchronization must therefore recover overwritten weights without a historical snapshot, translate changes across heterogeneous layouts without a complete checkpoint intermediate, and safely transfer variable-length updates to their destination shards.

To overcome these challenges, we present \textsc{AReaL-DTE}, (where \textsc{DTE} stands for \underline{D}elta \underline{T}ransfer \underline{E}ngine), a snapshot-free sparse weight-synchronization system designed for the data transfer between training and rollout inference micro services. \textsc{AReaL-DTE} organizes synchronization into three coordinated stages: reconstructing overwritten weights directly from live training state, detecting and remapping inference-visible changes across heterogeneous parameter layouts, and transferring the resulting receiver-ready sparse updates. The same sparse updates can be transferred directly within a cluster or through shared storage across clusters. By treating sparse synchronization as an end-to-end system problem, \textsc{AReaL-DTE} eliminates both the model-scale historical snapshot used for delta construction and the complete-model intermediate used to bridge training and inference layouts. Our contributions are summarized as follows:

\textbf{\underline{Contribution 1.} Verify and exploit the sparsity in RL weight updates.}
We verify that consecutive RL optimization steps modify only a small fraction of the weight values visible to rollout inference. Across Qwen3-8B and Qwen3-30B-A3B on GSM8K, HumanEval, MATH-500, and LogiQA, fewer than $2\%$ of \texttt{BF16} weight elements change between consecutive policy versions. Because rollout inference consumes \texttt{BF16} weights, elements whose \texttt{BF16} representations remain unchanged do not need to be transferred. This observation establishes a substantial opportunity to reduce network and shared-storage traffic. We further identify that realizing its end-to-end benefit requires eliminating the model-scale historical and intermediate states that remain in checkpoint-oriented delta synchronization.

\textbf{\underline{Contribution 2.}  Design and implement an optimized Delta Transfer Engine.}
We design and implement a set of coordinated optimizations that translate weight sparsity into latency and memory savings. \textsc{AReaL-DTE} analytically inverts each AdamW update to reconstruct overwritten weights on demand and streams reconstruction one converter unit at a time, avoiding a persistent historical model snapshot. It passes the reconstructed and current parameters through the same MCore-to-Hugging-Face converter and compares their \texttt{BF16} bit patterns in an aligned canonical space, restricting synchronization to changes represented in the inference data type. A precomputed transfer plan then remaps each changed element directly into receiver-local coordinates, avoiding complete-model reconstruction and resharding. 
For cross-cluster synchronization, \textsc{AReaL-DTE} publishes canonical sparse index--value chunks through shared storage, and the destination SGLang scheduler performs the final layout-aware sparse application after a committed version becomes visible.
Within a cluster, it first exchanges per-destination element counts---including zeros---and then transfers the corresponding indices and values through a deadlock-safe two-round protocol. Rollout workers directly scatter the received values into their inference shards. These optimizations jointly remove model-scale historical and intermediate states from delta construction, transfer, and application, while relying on the surrounding control and storage layer for version ordering and consistency semantics.

\textbf{\underline{Contribution 3.}  Demonstrate substantial improvements in comprehensive evaluations.}
We conduct a comprehensive evaluation of \textsc{AReaL-DTE} on Qwen3-8B and Qwen3-30B-A3B across GSM8K, HumanEval, MATH-500, and LogiQA. In storage-mediated cross-cluster experiments covering bandwidth limits from $0.1$ to $10$ Gbps, \textsc{AReaL-DTE} achieves speedups of up to $19.9\times$ over ByteCheckpoint and $3.2\times$ over \textsc{PULSE}. In the same-cluster setting, the corresponding speedups reach $7.6\times$ and $7.4\times$. On Qwen3-30B-A3B, its compressed sparse files occupy only $2.53$--$2.87$ GB, compared with approximately $44.5$ GB for dense checkpoints, and are $16.3\times$ smaller on average. \textsc{AReaL-DTE} also reduces same-cluster peak GPU memory by approximately $41\%$ and peak CPU memory by at least $87\%$. These results demonstrate that fully exploiting weight-update sparsity requires eliminating model-scale state not only from transfer, but also from update construction and application.

\section{Background and Related Work}
\label{sec:background}

\subsection{RL Post-Training Frameworks}

Policy-weight synchronization is a recurring systems interface in distributed RL post-training.
General-purpose frameworks such as HybridFlow, OpenRLHF, NeMo-Aligner, and ROLL coordinate rollout generation, reward computation, and policy optimization across specialized workers and parallel configurations~\cite{hybridflow2025,openrlhf2024,nemoaligner2024,roll2025}.
Each optimization cycle produces a new policy version that rollout workers must install before---or, in asynchronous pipelines, while---generating subsequent trajectories.
The synchronization path therefore affects both resource utilization and the freshness of rollout policies.
Recent systems increasingly decouple rollout generation from policy training.
AReaL overlaps the two stages while controlling rollout staleness, whereas AReaL-2.0, ProRL-Agent, RollArt, and FlexMARL expose finer-grained agent, rollout, environment, reward, and training services~\cite{fu2025areal,yan2026next,zhang2026prorl,gao2026rollart,jiang2026rollout}.
DORA maintains multiple policy versions for streaming asynchronous rollout, Laminar introduces relay workers as a distributed parameter service, and Weave co-schedules disaggregated rollout and training clusters~\cite{dora2026,laminar2025,weave2026}.
ROSE further recruits serving GPUs as elastic rollout capacity and includes shard- and sparsity-aware cross-cluster weight transfer~\cite{rose2026}.
These architectures improve scheduling flexibility and hardware specialization, but they also make frequent policy publication an explicit data-plane operation.
\sys complements this line of work by optimizing that operation rather than changing the surrounding rollout schedule, RL algorithm, or consistency policy.

\subsection{Model-State Transfer Across Heterogeneous Layouts}

Training and inference engines represent the same policy differently because they optimize for different workloads.
Megatron-LM and FSDP partition optimizer and model state for distributed training, whereas vLLM and SGLang organize weights for high-throughput serving~\cite{megatronlm,fsdp,vllm,sglang}.
Consequently, the two sides may use different parameter names, fusion and permutation rules, data types, and tensor-, pipeline-, data-, or expert-parallel layouts.
A synchronization mechanism must therefore translate policy state into the receiver's layout in addition to moving it. Full-state checkpoints provide a robust contract across these heterogeneous layouts.
ByteCheckpoint and Universal Checkpointing use parallelism-independent representations and load-time remapping to make distributed model state portable across configurations~\cite{wan2025bytecheckpoint,universalcheckpoint2025}; RL frameworks such as HybridFlow similarly support complete policy-state transfer between training and rollout workers~\cite{hybridflow2025}.
TensorHub takes a complementary approach: its reference-oriented storage abstraction locates live weight replicas and serves topology-optimized transfers without first copying those replicas into storage~\cite{tensorhub2026}.
For a policy with $N$ inference-visible elements, however, a dense refresh constructs, stages, transfers, and applies $O(N)$ values even when the new policy differs at only a small fraction of positions.
Checkpoint portability thus addresses a different objective from the high-frequency synchronization path studied here: \sys preserves layout translation while avoiding a complete-model representation for each refresh.

\subsection{Sparse Policy-Weight Synchronization}

Sparse synchronization reduces communication by publishing only inference-visible changes.
PULSE identifies that many FP32 optimizer updates do not alter the \texttt{BF16} values consumed by the next forward pass and transmits changed indices with their new \texttt{BF16} values~\cite{miahi2026understanding}.
SparrowRL packages lossless sparse deltas under inference-side parameter names and applies them directly to rollout workers connected through commodity and wide-area networks~\cite{ruan2026rl}.
ROSE retains previous-step weights, computes sparse arithmetic deltas, and partitions them by destination shard to avoid dense receiver-side materialization~\cite{rose2026}.
The slime framework likewise provides storage-mediated delta synchronization through canonical Hugging Face checkpoints~\cite{slime2026delta}.
If $M$ of $N$ elements change and $M\ll N$, these methods reduce the logical update payload from $O(N)$ values to $O(M)$ index--value pairs.
Payload sparsity alone, however, does not make the complete synchronization pipeline sparse.
PULSE retains a previous checkpoint for change detection, while ROSE retains the preceding weights and slime maintains canonical checkpoint copies for delta construction and application~\cite{miahi2026understanding,rose2026,slime2026delta}.
More generally, a checkpoint-oriented path may still gather and convert complete state before comparison, broadcast receiver-irrelevant entries, or reconstruct and reshard a complete representation at the destination.
\sys is distinguished by making all three local operations sparse: it reconstructs overwritten weights without a historical snapshot, performs change detection during training-to-inference conversion, and remaps changes into receiver-local coordinates before transfer.
It supports this receiver-ready representation through shared storage across clusters and direct point-to-point communication within a cluster.

\section{Key Observations and Design Rationale}
\label{sec:motivation}

We first present our key observations for RL training. Figure~\ref{fig:dte_motivation} contrasts a checkpoint-oriented synchronization path with the end-to-end sparse path in \sys.
Both paths begin with an updated training policy and must install the same inference-visible version on rollout workers.
The following observations identify why a sparse checkpoint can still incur dense local work and which information is already available to eliminate that work. We enumerate the key observations below: 

\begin{figure}[t]
    \centering
    \includegraphics[width=\columnwidth]{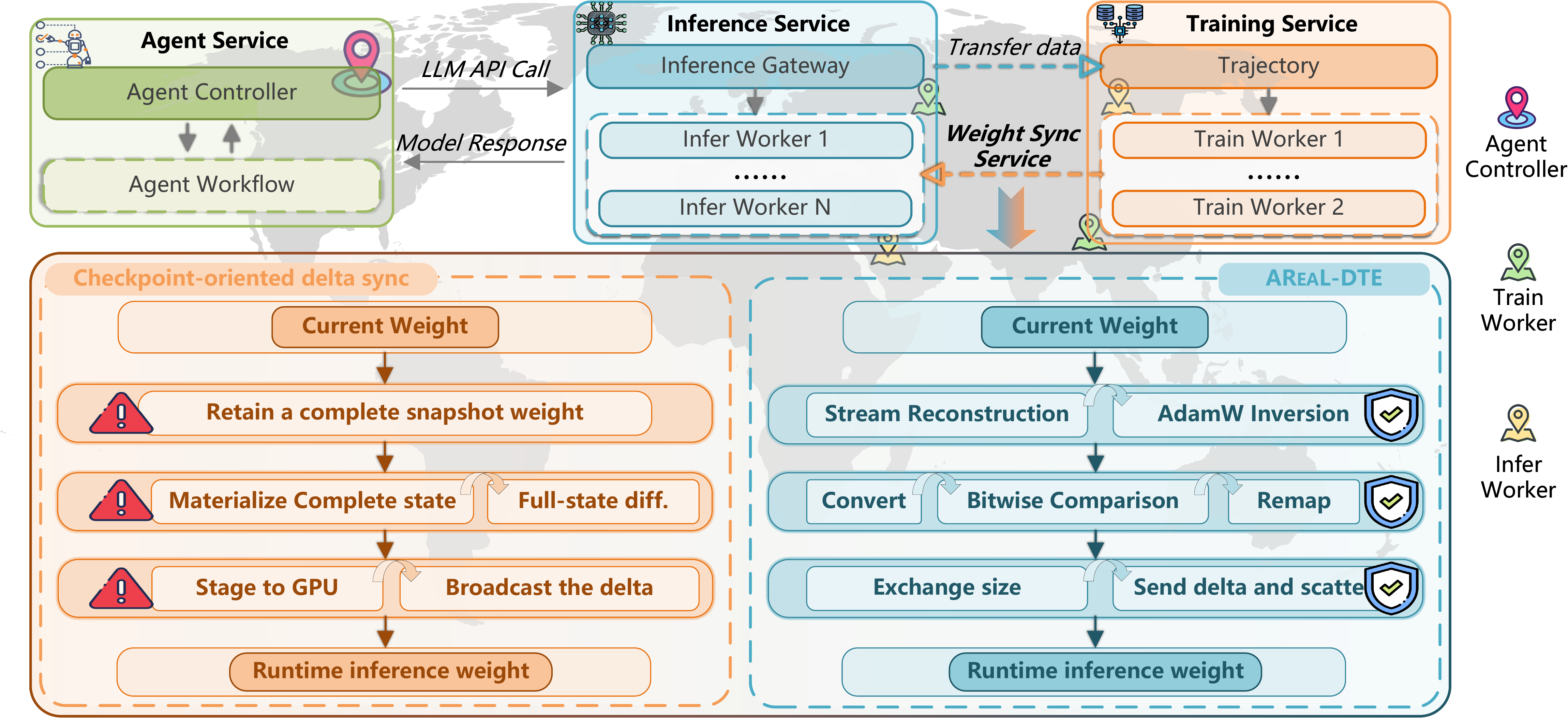}
    \caption{Motivation for \sys.
    Checkpoint-oriented delta synchronization may retain a historical model snapshot and materialize complete-model intermediates during update construction and application.
    \sys instead reconstructs previous weights through streaming AdamW inversion, detects and remaps changes during format conversion, and applies receiver-local updates directly to inference shards.}
    \label{fig:dte_motivation}
\end{figure}

\textbf{\underline{Observation 1:} policy changes are sparse in the receiver's data type.}
Across the evaluated workloads, fewer than $2\%$ of \texttt{BF16} weight elements differ between consecutive policy versions (Figure~\ref{fig:zstd_payload_comparison}(c)).
Because rollout inference consumes \texttt{BF16} weights, an FP32 optimizer update that leaves the \texttt{BF16} bit pattern unchanged has no effect on the rollout policy and need not be transferred.

\textbf{\underline{Observation 2:} the stated AdamW update can be inverted from live state.}
AdamW updates policy weights in place, so conventional differencing retains a complete previous policy.
In exact arithmetic, the current FP32 master weights, first- and second-moment states, optimizer step, and hyperparameters algebraically determine the pre-update weights under the update used by our training stack.
\sys can therefore compute a numerical reconstruction on demand instead of storing the historical policy persistently.

\textbf{\underline{Observation 3:} layout translation is deterministic even though layouts differ.}
Training and inference engines may fuse, permute, name, and shard parameters differently, but their converter and shard metadata define a stable mapping between the two layouts.
Applying the same converter to the previous and current versions creates an aligned logical space for comparison, and the overlap between that space and every receiver shard can be precomputed.

\textbf{\underline{Observation 4:} transfer topology is static while payload cardinality is dynamic.}
The training and inference layouts determine the sender--receiver pairs and their operation order before synchronization begins.
Only the number of changed elements assigned to each pair varies across policy versions, and that number may be zero.
This separation permits a fixed communication schedule even though buffer sizes are data dependent.

\textbf{Design intuitions.}
These observations lead to a receiver-oriented design: construct changes in the inference data type, translate them before communication, and avoid retaining or materializing state that can be derived on demand.
First, \sys inverts each AdamW update and streams one reconstructed converter unit at a time into change detection, eliminating the persistent historical snapshot.
Second, it converts the reconstructed and current parameters through the same MCore-to-Hugging-Face path, compares their \texttt{BF16} bit patterns, and immediately remaps changed indices into receiver-local coordinates.
This fused conversion, comparison, and remapping path avoids complete-model gathering, differencing, reconstruction, and resharding.
Third, \sys separates size discovery from data movement: it exchanges a count for every planned operation, including zero-length operations, before transferring receiver-specific indices and values.
Each rollout worker can then scatter the received values directly into its live inference shards without a generic-delta broadcast or receiver-side layout conversion.

\section{\sys Design and Implementation}
\label{sec:method}

\begin{figure*}[t]
    \centering
    \includegraphics[width=0.9\textwidth]{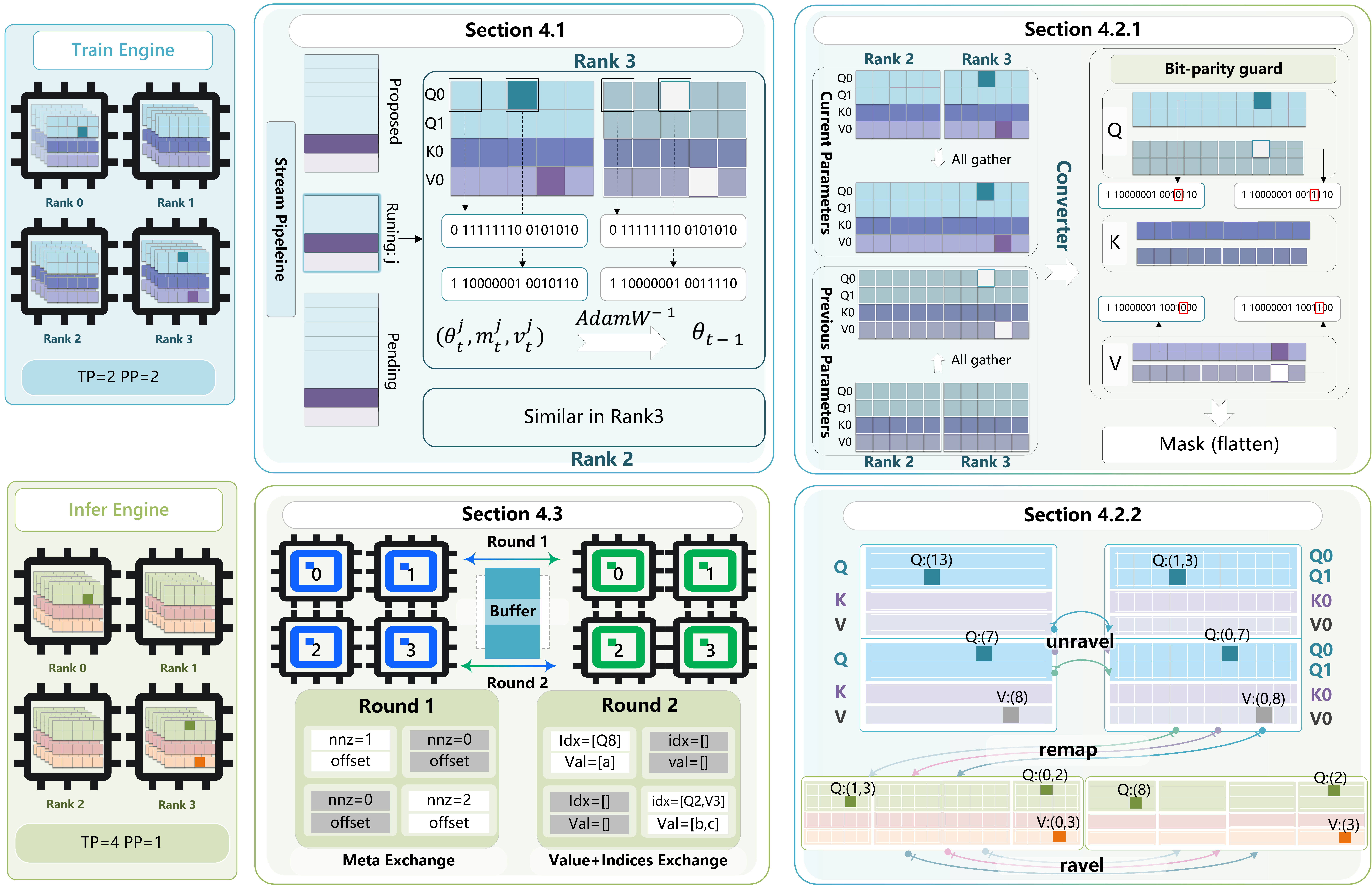}
    \caption{End-to-end same-cluster pipeline of \sys.
    Starting from optimizer-native Megatron Core (MCore) shards, \sys
    (1) reconstructs previous parameters through streaming AdamW inversion,
    (2) performs converter-aligned \texttt{BF16} change detection and maps
    canonical changed indices into receiver-local coordinates through
    unravel--remap--ravel operations, and
    (3) delivers receiver-local sparse updates for direct application.}
    \label{fig:overview}
\end{figure*}

We introduce the design and implementation of \sys. Figure~\ref{fig:overview} shows how \sys transforms optimizer-native training state into receiver-local sparse updates.
The pipeline proceeds clockwise through three logical stages.
Section~\ref{sec:optimizer_inversion} streams over live M-Core parameters and reconstructs each pre-update parameter from the post-update weight and retained AdamW state.
Sections~\ref{sec:canonical_delta_detection} and~\ref{sec:receiver_space_remapping}
align the reconstructed and current versions in a canonical Hugging Face
representation, compare their \texttt{BF16} bit patterns, and, for the
same-cluster path, map the changed indices into receiver-local coordinates.
Finally, \sys delivers receiver-local sparse payloads through the same-cluster
protocol in Section~\ref{sec:two_round_transfer}; for the cross-cluster path in
Section~\ref{sec:cross_cluster_transfer}, it publishes canonical sparse changes
through shared storage for final layout-aware application by the destination
SGLang scheduler.

\subsection{Optimizer-Inverted Weight Reconstruction}
\label{sec:optimizer_inversion}

The reconstruction stage removes the need to retain a complete pre-update policy.
\sys targets AdamW~\cite{loshchilov2019decoupled}, which updates the FP32 master weights in place, and uses the retained optimizer state to reconstruct the overwritten values.

Let $\theta_t$ denote an FP32 master-weight element after optimizer step $t$, let $g_t$ denote its gradient, and let $m_t$ and $v_t$ denote the corresponding first- and second-moment states.
The moment states are updated as
\begin{equation}
    m_t
    =
    \beta_1 m_{t-1}
    +
    (1-\beta_1)g_t,
    \qquad
    v_t
    =
    \beta_2 v_{t-1}
    +
    (1-\beta_2)g_t^2,
    \label{eq:adamw_moments}
\end{equation}
where $\beta_1$ and $\beta_2$ are the moment coefficients and $g_t^2$ denotes element-wise squaring.
All expressions apply element-wise to a master-weight tensor.

Under the decoupled AdamW update used by our training stack, the predecessor $\theta_{t-1}$ satisfies an algebraic inverse.
Let $\eta_t$, $\lambda$, and $\epsilon$ denote the learning rate, parameter-group weight-decay coefficient, and numerical-stability constant, respectively.
\sys evaluates the inverse as
\begin{equation}
    \widehat{\theta}_{t-1}
    =
    \frac{
        \theta_t
        +
        \eta_t
        \dfrac{
            m_t/(1-\beta_1^t)
        }{
            \sqrt{v_t/(1-\beta_2^t)}+\epsilon
        }
    }{
        1-\eta_t\lambda
    }.
    \label{eq:adamw_inverse}
\end{equation}

All quantities on the right-hand side remain available after the optimizer step, allowing \sys to compute $\widehat{\theta}_{t-1}$ without storing $\theta_{t-1}$; Appendix~\ref{app:adamw_derivation} provides the derivation.
Equation~\ref{eq:adamw_inverse} is exact for the stated update in exact arithmetic.
Because fused finite-precision implementations can round intermediate results, $\widehat{\theta}_{t-1}$ denotes the numerical reconstruction used by the change detector rather than a general claim of bitwise recovery of the FP32 master weight.

\sys applies Equation~\ref{eq:adamw_inverse} in the optimizer-native MCore space, where every FP32 master-weight shard is element-wise aligned with its moment states.
Each data-parallel rank reconstructs its locally owned optimizer slice, and ranks in the optimizer group gather the reconstructed slices required for the corresponding tensor-parallel (TP) MCore parameter.

Streaming bounds the transient memory required by reconstruction.
Rather than materializing all reconstructed parameters simultaneously, \sys reconstructs one converter unit, consumes it during conversion and change detection, and then releases the temporary tensor.
A converter unit is one MCore parameter or the smallest group that its conversion rule must process together.
This pipeline eliminates the persistent historical snapshot and bounds transient reconstruction state by the largest converter unit rather than the full model.

\subsection{Converter-Aligned Delta Detection and Remapping}
\label{sec:converter_aligned_remapping}

The second stage makes sparse changes meaningful across heterogeneous layouts.
It first aligns the reconstructed and current versions in a canonical parameter space and then remaps the detected changes into receiver-local coordinates.

\subsubsection{Canonical-Space Delta Detection.}
\label{sec:canonical_delta_detection}

Canonical comparison is necessary because a training-side index cannot be applied directly to an inference tensor without accounting for fusion, permutation, naming, and sharding.
Appendix~\ref{app:mcore_hf_layouts} illustrates this mismatch for a fused QKV parameter.
\sys passes the reconstructed and current versions from sender group $s$ through the same MCore-to-Hugging-Face converter $\mathcal{C}_s$.

For each canonical Hugging Face parameter $p$, let $W_{t-1,s}^{p}$ and $W_{t,s}^{p}$ denote the converted previous and current tensors emitted by sender group $s$, respectively:
\begin{equation}
    \left\{
        (p,W_{t-1,s}^{p})
    \right\}
    =
    \mathcal{C}_s(\widehat{\theta}_{t-1}),
    \qquad
    \left\{
        (p,W_{t,s}^{p})
    \right\}
    =
    \mathcal{C}_s(\theta_t).
    \label{eq:aligned_conversion}
\end{equation}

The converter performs the tensor-local TP or expert-parallel (EP) gathering, fused-tensor splitting, permutation, and parameter renaming required to place both versions in the same organization and element order.
For example, it gathers and separates the fused QKV parameter in Figure~\ref{fig:overview} into canonical Q, K, and V tensors before comparison.

\sys casts both tensors to \texttt{BF16}, the data type consumed by the inference engine, and denotes the resulting tensors as
$\overline{W}_{\tau,s}^{p}
=
\mathcal{Q}_{\mathrm{\texttt{BF16}}}(W_{\tau,s}^{p})$
for $\tau\in\{t-1,t\}$.
It then reinterprets the \texttt{BF16} tensors as 16-bit integer views, compares their stored bit patterns element-wise, and extracts the changed indices and target values:
\begin{align}
    M_{t,s}^p
    &=
    \operatorname{vec}
    \left(
        \mathcal{B}_{16}(\overline{W}_{t,s}^p)
        \neq
        \mathcal{B}_{16}(\overline{W}_{t-1,s}^p)
    \right),
    \nonumber\\
    I_{t,s}^p
    &=
    \operatorname{nonzero}(M_{t,s}^p),
    \qquad
    V_{t,s}^p
    =
    \operatorname{vec}(\overline{W}_{t,s}^p)[I_{t,s}^p].
    \label{eq:canonical_delta_detection}
\end{align}

Here, $\mathcal{B}_{16}(\cdot)$ denotes bit-preserving reinterpretation, $\operatorname{vec}(\cdot)$ flattens a tensor, $I_{t,s}^p$ contains the changed sender-local canonical indices, and $V_{t,s}^p$ contains their target \texttt{BF16} values.

Comparing stored \texttt{BF16} bit patterns restricts the payload to differences represented in the receiver's data type.
Sending target values makes application idempotent for a fixed version and avoids error accumulation from repeatedly adding quantized arithmetic differences.
Because conversion and comparison operate on one converter unit at a time, the required TP or EP gathers remain tensor-local and do not materialize a complete model state.

\subsubsection{Receiver-Space Delta Remapping.}
\label{sec:receiver_space_remapping}

Receiver-space remapping turns each canonical change into an index that the destination can apply directly.
When training and inference use different layouts, a canonical parameter region may be partitioned, replicated, or assigned to different ranks on the two sides.
\sys therefore precomputes a transfer plan $\mathcal{T}$ from the training and inference shard metadata.

Each operation $o\in\mathcal{T}$ identifies a canonical parameter $p$, sender $s$, receiver $r$, and the hyperrectangular overlap between their tensor views.
Let $\mathbf{n}_s^p$ and $\mathbf{n}_r^p$ denote the shapes of the sender-local converted tensor and receiver-local shard.
Let $\mathbf{a}_s^o$ and $\mathbf{a}_r^o$ denote the corresponding overlap origins, and let $\mathbf{e}^o$ denote the overlap extent.
For each flat index $i\in I_{t,s}^p$, \sys first recovers its multidimensional sender coordinate:
\begin{equation}
    \mathbf{u}
    =
    \operatorname{unravel}
    \left(
        i,\mathbf{n}_s^p
    \right).
    \label{eq:canonical_unravel}
\end{equation}

The change belongs to operation $o$ when
\(
\mathbf{a}_s^o \leq \mathbf{u} <
\mathbf{a}_s^o+\mathbf{e}^o
\)
component-wise.
\sys translates each matching coordinate into the receiver system and flattens it using the receiver shard shape:
\begin{equation}
    j_r
    =
    \operatorname{ravel}
    \left(
        \mathbf{u}
        -
        \mathbf{a}_s^o
        +
        \mathbf{a}_r^o,
        \mathbf{n}_r^p
    \right).
    \label{eq:receiver_local_remapping}
\end{equation}

Remapping preserves the associated \texttt{BF16} target value and produces a receiver-specific entry $(r,j_r,v)$.
Applying the procedure to every operation in $\mathcal{T}$ partitions or replicates changes according to the inference layout and produces an ordered sparse payload for each sender--receiver pair.
The resulting indices can be applied to destination shards without reconstructing or resharding a complete model.
\begin{table*}[t]
    \centering
    \caption{Same-cluster results on \textbf{Qwen3-30B-A3B}.}
    \label{tab:qwen3_30b_same_cluster}

    \setlength{\tabcolsep}{3pt}
    \renewcommand{\arraystretch}{1.12}
    \footnotesize

    \begin{tabular}{
        @{}
        >{\centering\arraybackslash}p{2.10cm}
        >{\centering\arraybackslash}p{2.10cm}
        >{\centering\arraybackslash}p{2.00cm}
        @{\hspace{5pt}}
        *{2}{>{\centering\arraybackslash}p{1.80cm}}
        @{\hspace{5pt}}
        *{2}{>{\centering\arraybackslash}p{1.80cm}}
        @{}
    }
        \toprule

        \rowcolor{white}
        & & & \multicolumn{2}{c}{\textbf{ByteCheckpoint}} & \multicolumn{2}{c}{\textbf{PULSE}} \\

        \rowcolor{subheaderGray}
        \cellcolor{white}\multirow{-2}{*}{\textbf{Dataset}} &
        \cellcolor{white}\multirow{-2}{*}{\textbf{Metric}} &
        \cellcolor{white}\multirow{-2}{*}{\mbox{\textbf{\textcolor{oursAccent}{\sys}}}} &
        \textbf{NCCL} & \textbf{Disk} & \textbf{NCCL} & \textbf{Disk} \\

        \midrule

        \rowcolor{gsmLight}
        \cellcolor{gsmDark!14} & E2E time $\downarrow$ & \cellcolor{oursRow}\textbf{11.58\thinspace s} & 88.03\thinspace s & 226.53\thinspace s & 59.42\thinspace s & 93.40\thinspace s \\
        \rowcolor{gsmLight}
        \cellcolor{gsmDark!14} & GPU $\downarrow$ & \cellcolor{oursRow}\textbf{71.78\thinspace GB} & 121.82\thinspace GB & 121.28\thinspace GB & 129.17\thinspace GB & 129.15\thinspace GB \\
        \rowcolor{gsmLight}
        \cellcolor{gsmDark!14}\multirow{-3}{*}{\textbf{\textcolor{gsmDark}{GSM8K}}} & CPU $\downarrow$ & \cellcolor{oursRow}\textbf{15.33\thinspace GB} & 129.29\thinspace GB & 125.53\thinspace GB & 141.28\thinspace GB & 139.19\thinspace GB \\

        \addlinespace[3pt]

        \rowcolor{humanLight}
        \cellcolor{humanDark!14} & E2E time $\downarrow$ & \cellcolor{oursRow}\textbf{11.83\thinspace s} & 83.43\thinspace s & 227.21\thinspace s & 64.80\thinspace s & 101.16\thinspace s \\
        \rowcolor{humanLight}
        \cellcolor{humanDark!14} & GPU $\downarrow$ & \cellcolor{oursRow}\textbf{71.45\thinspace GB} & 120.12\thinspace GB & 121.20\thinspace GB & 130.18\thinspace GB & 129.20\thinspace GB \\
        \rowcolor{humanLight}
        \cellcolor{humanDark!14}\multirow{-3}{*}{\textbf{\textcolor{humanDark}{HumanEval}}} & CPU $\downarrow$ & \cellcolor{oursRow}\textbf{15.45\thinspace GB} & 126.65\thinspace GB & 123.17\thinspace GB & 139.68\thinspace GB & 138.64\thinspace GB \\

        \addlinespace[3pt]

        \rowcolor{mathLight}
        \cellcolor{mathDark!14} & E2E time $\downarrow$ & \cellcolor{oursRow}\textbf{12.23\thinspace s} & 82.67\thinspace s & 227.59\thinspace s & 65.66\thinspace s & 96.46\thinspace s \\
        \rowcolor{mathLight}
        \cellcolor{mathDark!14} & GPU $\downarrow$ & \cellcolor{oursRow}\textbf{71.34\thinspace GB} & 122.03\thinspace GB & 121.55\thinspace GB & 129.01\thinspace GB & 129.10\thinspace GB \\
        \rowcolor{mathLight}
        \cellcolor{mathDark!14}\multirow{-3}{*}{\textbf{\textcolor{mathDark}{MATH-500}}} & CPU $\downarrow$ & \cellcolor{oursRow}\textbf{15.44\thinspace GB} & 123.21\thinspace GB & 122.79\thinspace GB & 137.04\thinspace GB & 139.83\thinspace GB \\

        \addlinespace[3pt]

        \rowcolor{logiqaLight}
        \cellcolor{logiqaDark!14} & E2E time $\downarrow$ & \cellcolor{oursRow}\textbf{12.14\thinspace s} & 89.48\thinspace s & 228.27\thinspace s & 62.00\thinspace s & 92.14\thinspace s \\
        \rowcolor{logiqaLight}
        \cellcolor{logiqaDark!14} & GPU $\downarrow$ & \cellcolor{oursRow}\textbf{71.19\thinspace GB} & 121.95\thinspace GB & 121.87\thinspace GB & 128.57\thinspace GB & 128.29\thinspace GB \\
        \rowcolor{logiqaLight}
        \cellcolor{logiqaDark!14}\multirow{-3}{*}{\textbf{\textcolor{logiqaDark}{LogiQA}}} & CPU $\downarrow$ & \cellcolor{oursRow}\textbf{15.40\thinspace GB} & 123.80\thinspace GB & 121.49\thinspace GB & 141.60\thinspace GB & 140.50\thinspace GB \\

        \bottomrule
    \end{tabular}

    \vspace{3pt}

    \parbox{\linewidth}{%
        \footnotesize
        \textit{Notes.}
        E2E denotes end-to-end synchronization latency.
        Memory consumption is reported in decimal GB.
        Lower values are better.
    }
\end{table*}
\subsection{Deadlock-Safe Two-Round Delta Transfer}
\label{sec:two_round_transfer}

The same-cluster path must transfer variable-length payloads without misaligning point-to-point (P2P) calls.
After remapping, each operation $o\in\mathcal{T}$ produces $(I_o,V_o)$, where $I_o$ contains receiver-local flat indices and $V_o$ contains the corresponding \texttt{BF16} target values.
The entry count $c_o=|I_o|$ varies by policy version and may be zero even though the transfer plan fixes the sender, receiver, and operation order.
The receiver cannot allocate exact buffers until it learns $c_o$, and independently skipping an empty operation on only one side would shift the ordered P2P schedule and could deadlock later operations.

\sys resolves both issues in two rounds.
In the first round, each sender transmits $c_o$ for every operation in transfer-plan order, including $c_o=0$, and each receiver allocates buffers of the required size.
In the second round, senders transmit $I_o$ and $V_o$ in the same order.
For $c_o=0$, both peers retain the operation's position and issue matched zero-count calls with empty buffers; subsequent calls therefore remain aligned.

For each nonempty payload, receiver $r$ directly scatters the received values into the live inference shard of parameter $p$:
\begin{equation}
    \operatorname{vec}(W_r^p)[I_o]
    \leftarrow
    V_o.
    \label{eq:direct_sparse_apply}
\end{equation}

Because the received indices are already receiver-local, application requires neither complete-model reconstruction nor an additional resharding pass.
After all planned payloads for a policy version have been applied, \sys reports completion to the surrounding control plane.
That control plane retains responsibility for version ordering and for making the completed policy visible to new rollouts, so a partially applied version is never reported as synchronized.

\subsection{Storage-Mediated Cross-Cluster Transfer}
\label{sec:cross_cluster_transfer}




Section~\ref{sec:two_round_transfer} delivers receiver-local sparse payloads through direct P2P communication when the training and inference engines share a communication group. Across clusters, such a group is unavailable. \sys therefore transports the canonical BF16 sparse changes produced by Section~\ref{sec:canonical_delta_detection} through an HTTP-backed shared filesystem. Unlike the same-cluster path, which remaps indices before transfer, the storage-mediated path performs the final inference-layout mapping inside the destination SGLang scheduler.

On the training side, \sys stages gathered canonical tensors through a bounded pinned-memory pipeline and serializes only changed index--value pairs into compressed payload chunks. This design bounds staging memory and avoids materializing a complete canonical checkpoint before publication.

To publish one version consistently, \sys designates writer ranks responsible for the payload partitions assigned to them; not every training rank is necessarily a writer. All participating ranks coordinate their publication metadata, including ranks whose assigned partitions contain no changes. A designated coordinator commits a version manifest only after this coordination completes. The manifest records the version chain, payload locations, lengths, and checksums, allowing the destination to locate the required chunks without
the explicit count-exchange round used by the same-cluster P2P protocol. If no writer has a changed payload, \sys publishes no new version and retains the previous synchronization base, preventing sender--receiver version-chain divergence.

After the destination service retrieves a committed version from shared storage, its local SGLang scheduler verifies that the delta extends its locally applied base version and selectively applies the sparse entries to the live inference model. When a mapping is known, \sys directly scatters entries into the corresponding inference shard; otherwise, it invokes the inference engine's standard loader while preserving unchanged parameter positions. Thus, the storage-mediated path avoids reconstructing a complete HuggingFace checkpoint or reloading all model weights for each policy version.
\begin{figure*}[t]
    \centering
    \includegraphics[
        width=\textwidth
    ]{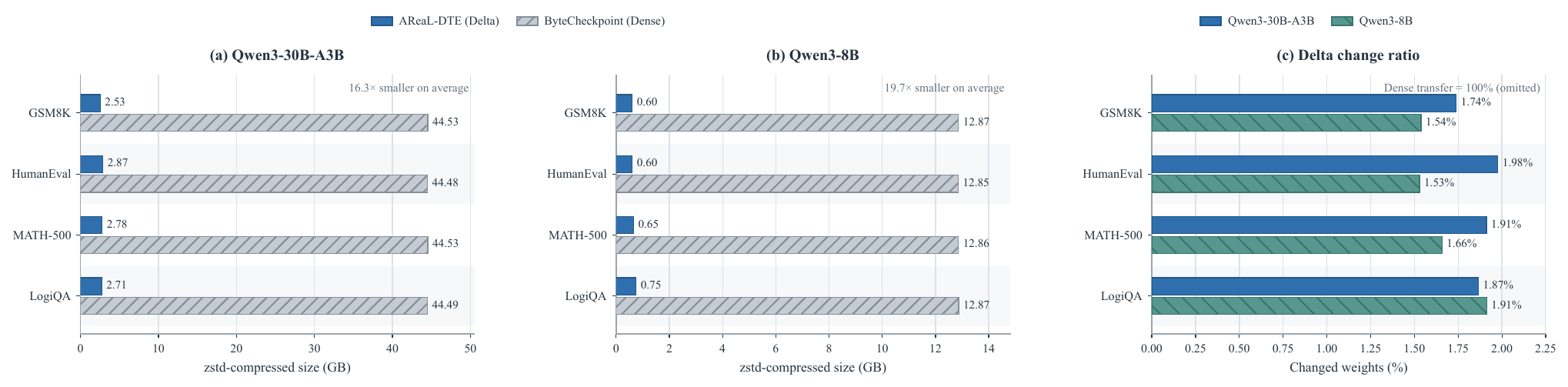}
    \caption{
        Zstd-compressed synchronization-artifact size and changed-weight fraction.
        Panels (a) and (b) compare the compressed artifacts transmitted by
        \sys and ByteCheckpoint on Qwen3-30B-A3B and Qwen3-8B,
        respectively.
        Panel (c) reports the fraction of changed weights transmitted by
        \sys; the $100\%$ dense-transfer fraction is omitted.
    }
    \label{fig:zstd_payload_comparison}
\end{figure*}
\section{Evaluation}
\label{sec:evaluation}

We organize the evaluation to answer the following three key research questions:
\begin{enumerate}[leftmargin=*,label=\textbf{RQ\arabic*:},itemsep=2pt,topsep=3pt]
    \item \textbf{Same-cluster efficiency.} How much does \sys reduce end-to-end synchronization latency and peak GPU and CPU memory when training and rollout workers run in the same cluster (Section~\ref{sec:same_cluster_results})?
    \item \textbf{Cross-cluster efficiency.} How do inter-cluster bandwidth and transferred data volume affect the latency and training-loop overhead of \sys relative to the baselines (Section~\ref{sec:cross_cluster_results})?
    \item \textbf{Generality across model scales.} Do the latency, transfer-volume, and memory benefits persist on the smaller dense Qwen3-8B model as well as the larger Qwen3-30B-A3B mixture-of-experts model (Section~\ref{sec:qwen3_8b_results})?
\end{enumerate}

\subsection{Experimental Setup}
\label{sec:experimental_setup}

\textbf{Models and workloads.}
Our evaluation covers both dense and mixture-of-experts architectures at two scales: Qwen3-8B and Qwen3-30B-A3B.
For each model, we run online RL workloads on GSM8K, HumanEval, MATH-500, and LogiQA.
We collect every synchronization input after an actual policy-optimization step so that the measured change fractions and payloads reflect training-generated updates.

\textbf{Baselines.}
We compare \sys against ByteCheckpoint and PULSE as representative dense and sparse synchronization baselines, respectively.
ByteCheckpoint transfers a complete model state at each refresh.
PULSE transfers changed \texttt{BF16} values but retains the previous checkpoint used to construct the patch.
Within a cluster, we evaluate direct NVIDIA Collective Communications Library (NCCL) and disk-mediated variants of both baselines; \sys uses its direct sparse point-to-point path.
Across clusters, all methods use their storage-mediated paths because the training and rollout workers do not share an NCCL communication group.
\begin{figure*}[t]
    \centering
    \includegraphics[width=\textwidth]{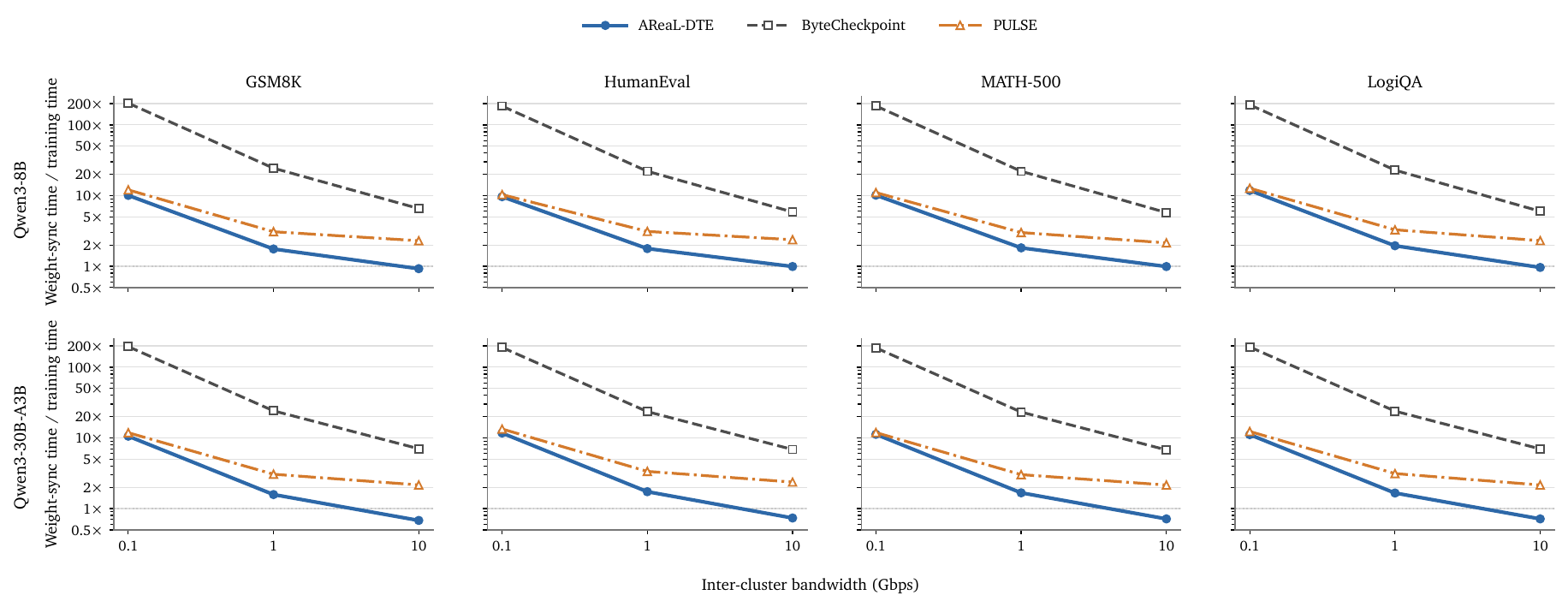}
    \caption{
        Cross-cluster weight-synchronization overhead under different inter-cluster bandwidths.
        The overhead is normalized by the training-step time excluding weight synchronization,
        i.e., $T_{\mathrm{weight\ sync}} / T_{\mathrm{step-without-weight}}$.
        The $y$-axis uses a logarithmic scale, and the horizontal $1\times$ line indicates that
        weight synchronization takes the same amount of time as the remaining training step.
    }
    \label{fig:cross_cluster_weight_overhead}
\end{figure*}

\textbf{Evaluation settings.}
All experiments run on NVIDIA H200 GPUs; Qwen3-8B and Qwen3-30B-A3B use 8 and 16 GPUs, respectively.
In the same-cluster setting, the training and inference engines communicate within one high-bandwidth cluster.
The cross-cluster analysis considers effective inter-cluster bandwidths of $0.1$, $1$, and $10$ Gbps, spanning constrained through relatively well-provisioned links.
Each method compresses its storage-mediated artifact with Zstandard (Zstd), transfers it through Alibaba Cloud Object Storage Service (OSS), and applies the downloaded update at the destination.
For bandwidth $B$, the reported cross-cluster latency is derived as
\(
T_{\mathrm{same,disk}}+
T_{\mathrm{upload,p50}}(B)+
T_{\mathrm{download,p50}}(B)
\),
where \(T_{\mathrm{same,disk}}\) includes local construction and receiver-side application and the remaining terms are median storage-transfer times.

\textbf{Metrics.}
We report end-to-end (E2E) synchronization latency and peak GPU and CPU memory consumption.
For same-cluster runs, E2E latency spans update construction, communication, and application.
For cross-cluster runs, it is the derived quantity defined above and includes the corresponding storage upload and download terms.
Peak memory is the maximum GPU or CPU memory observed during update construction, staging, communication, and receiver-side application.

\subsection{Same-Cluster Performance}
\label{sec:same_cluster_results}

\textbf{Answer to RQ1: \sys substantially reduces both synchronization latency and peak memory within a cluster.}
Tables~\ref{tab:qwen3_8b_same_cluster} and~\ref{tab:qwen3_30b_same_cluster} report the results.
Because no constrained inter-cluster link dominates these runs, they isolate the local costs of constructing, communicating, and applying each update.

On Qwen3-30B-A3B, \sys completes synchronization in $11.58$--$12.23$ seconds, compared with $82.67$--$89.48$ seconds for ByteCheckpoint-NCCL and $59.42$--$65.66$ seconds for PULSE-NCCL.
\sys therefore achieves a $6.8$--$7.6\times$ speedup over ByteCheckpoint-NCCL and a $5.1$--$5.5\times$ speedup over PULSE-NCCL.
The storage and staging work in the disk-mediated baseline variants makes them consistently slower than their direct NCCL variants.

\sys also limits peak GPU memory to $71.19$--$71.78$ GB and peak CPU memory to $15.33$--$15.45$ GB on Qwen3-30B-A3B.
Relative to the lowest-memory baseline configuration for each workload, these values represent an approximately $41\%$ GPU-memory reduction and at least an $87\%$ CPU-memory reduction.
On Qwen3-8B, \sys completes synchronization in $2.77$--$2.87$ seconds and reaches speedups of up to $4.0\times$ over ByteCheckpoint-NCCL and $7.4\times$ over PULSE-NCCL; Section~\ref{sec:qwen3_8b_results} analyzes the cross-model trend.

\begin{table*}[t]
    \centering

    \caption{Cross-cluster weight-synchronization results on \textbf{Qwen3-30B-A3B}. Lower values are better.}
    \label{tab:qwen3_30b_cross_cluster}

    \setlength{\tabcolsep}{2.5pt}
    \renewcommand{\arraystretch}{1.12}
    \scriptsize
    \normalfont

    {\bfseries (a) Derived E2E synchronization latency}
    \par\vspace{1pt}

    \begin{tabular}{
        @{}
        >{\centering\arraybackslash}p{1.75cm}
        *{3}{>{\centering\arraybackslash}p{1.35cm}}
        @{\hspace{6pt}}
        *{3}{>{\centering\arraybackslash}p{1.35cm}}
        @{\hspace{6pt}}
        *{3}{>{\centering\arraybackslash}p{1.35cm}}
        @{}
    }
        \toprule

        \rowcolor{white}
        & \multicolumn{3}{c}{\textbf{\textcolor{oursAccent}{\sys}}}
        & \multicolumn{3}{c}{ByteCheckpoint}
        & \multicolumn{3}{c}{PULSE} \\
        
        \rowcolor{subheaderGray}
        \cellcolor{white}\multirow{-2}{*}{\textbf{Dataset}}
        & \textbf{0.1 Gbps} & \textbf{1 Gbps} & \textbf{10 Gbps}
        & \textbf{0.1 Gbps} & \textbf{1 Gbps} & \textbf{10 Gbps}
        & \textbf{0.1 Gbps} & \textbf{1 Gbps} & \textbf{10 Gbps} \\

        \midrule

        \rowcolor{gsmLight}
        \cellcolor{gsmDark!14}\textbf{\textcolor{gsmDark}{GSM8K}} & \cellcolor{oursRow}477.47\thinspace s & \cellcolor{oursRow}71.31\thinspace s & \cellcolor{oursRow}30.70\thinspace s & 8741.18\thinspace s & 1080.12\thinspace s & 314.05\thinspace s & 544.86\thinspace s & 140.90\thinspace s & 99.43\thinspace s \\

        \rowcolor{humanLight}
        \cellcolor{humanDark!14}\textbf{\textcolor{humanDark}{HumanEval}} & \cellcolor{oursRow}539.64\thinspace s & \cellcolor{oursRow}79.88\thinspace s & \cellcolor{oursRow}33.91\thinspace s & 8741.01\thinspace s & 1080.69\thinspace s & 314.73\thinspace s & 604.60\thinspace s & 152.01\thinspace s & 107.34\thinspace s \\

        \rowcolor{mathLight}
        \cellcolor{mathDark!14}\textbf{\textcolor{mathDark}{MATH-500}} & \cellcolor{oursRow}523.59\thinspace s & \cellcolor{oursRow}77.88\thinspace s & \cellcolor{oursRow}33.32\thinspace s & 8740.88\thinspace s & 1081.11\thinspace s & 315.14\thinspace s & 567.42\thinspace s & 144.20\thinspace s & 102.75\thinspace s \\

        \rowcolor{logiqaLight}
        \cellcolor{logiqaDark!14}\textbf{\textcolor{logiqaDark}{LogiQA}} & \cellcolor{oursRow}511.81\thinspace s & \cellcolor{oursRow}76.45\thinspace s & \cellcolor{oursRow}32.92\thinspace s & 8744.39\thinspace s & 1082.10\thinspace s & 315.85\thinspace s & 559.26\thinspace s & 141.93\thinspace s & 97.78\thinspace s \\

        \bottomrule
    \end{tabular}

    \par\vspace{-1pt}

    {\bfseries (b) Peak memory consumption}
    \par\vspace{1pt}

    \begin{tabular}{
        @{}
        >{\centering\arraybackslash}p{1.75cm}
        *{2}{>{\centering\arraybackslash}p{2.15cm}}
        @{\hspace{6pt}}
        *{2}{>{\centering\arraybackslash}p{2.15cm}}
        @{\hspace{6pt}}
        *{2}{>{\centering\arraybackslash}p{2.15cm}}
        @{}
    }
        \toprule

        \rowcolor{white}
        & \multicolumn{2}{c}{\textbf{\textcolor{oursAccent}{\sys}}}
        & \multicolumn{2}{c}{ByteCheckpoint}
        & \multicolumn{2}{c}{PULSE} \\
        
        \rowcolor{subheaderGray}
        \cellcolor{white}\multirow{-2}{*}{\textbf{Dataset}}
        & \textbf{GPU} & \textbf{CPU}
        & \textbf{GPU} & \textbf{CPU}
        & \textbf{GPU} & \textbf{CPU} \\

        \midrule

        \rowcolor{gsmLight}
        \cellcolor{gsmDark!14}\textbf{\textcolor{gsmDark}{GSM8K}} & \cellcolor{oursRow}71.52\thinspace GB & \cellcolor{oursRow}40.56\thinspace GB & 121.28\thinspace GB & 125.53\thinspace GB & 129.15\thinspace GB & 139.19\thinspace GB \\

        \rowcolor{humanLight}
        \cellcolor{humanDark!14}\textbf{\textcolor{humanDark}{HumanEval}} & \cellcolor{oursRow}71.96\thinspace GB & \cellcolor{oursRow}37.58\thinspace GB & 121.20\thinspace GB & 123.17\thinspace GB & 129.20\thinspace GB & 138.64\thinspace GB \\

        \rowcolor{mathLight}
        \cellcolor{mathDark!14}\textbf{\textcolor{mathDark}{MATH-500}} & \cellcolor{oursRow}71.84\thinspace GB & \cellcolor{oursRow}37.44\thinspace GB & 121.55\thinspace GB & 122.79\thinspace GB & 129.10\thinspace GB & 139.83\thinspace GB \\

        \rowcolor{logiqaLight}
        \cellcolor{logiqaDark!14}\textbf{\textcolor{logiqaDark}{LogiQA}} & \cellcolor{oursRow}71.88\thinspace GB & \cellcolor{oursRow}40.75\thinspace GB & 121.87\thinspace GB & 121.49\thinspace GB & 128.29\thinspace GB & 140.50\thinspace GB \\

        \bottomrule
    \end{tabular}

    \par\vspace{2pt}

    \parbox{\linewidth}{%
        \scriptsize
        \raggedright
        \textit{Notes.}
        All methods use their disk-mediated transfer paths.
        Cross-cluster latency is derived as
        \(T_{\mathrm{same,disk}} + T_{\mathrm{upload,p50}} + T_{\mathrm{download,p50}}\).
        Memory values are taken from the corresponding same-cluster disk runs and reported in decimal GB.
    }
\end{table*}

\subsection{Cross-Cluster Performance}
\label{sec:cross_cluster_results}

\textbf{Answer to RQ2: \sys preserves its latency advantage across bandwidth regimes because its sparse artifacts reduce both transferred data and local processing.}
Tables~\ref{tab:qwen3_30b_cross_cluster} and~\ref{tab:qwen3_8b_cross_cluster} report the derived cross-cluster latency and memory results at all three bandwidths.

\begin{table*}[t]
    \centering
    \caption{Same-cluster results on \textbf{Qwen3-8B}.}
    \label{tab:qwen3_8b_same_cluster}

    \setlength{\tabcolsep}{3pt}
    \renewcommand{\arraystretch}{1.12}
    \footnotesize

    \begin{tabular}{
        @{}
        >{\centering\arraybackslash}p{2.10cm}
        >{\centering\arraybackslash}p{2.10cm}
        >{\centering\arraybackslash}p{2.00cm}
        @{\hspace{5pt}}
        *{2}{>{\centering\arraybackslash}p{1.80cm}}
        @{\hspace{5pt}}
        *{2}{>{\centering\arraybackslash}p{1.80cm}}
        @{}
    }
        \toprule

        \rowcolor{white}
        & & & \multicolumn{2}{c}{\textbf{ByteCheckpoint}} & \multicolumn{2}{c}{\textbf{PULSE}} \\

        \rowcolor{subheaderGray}
        \cellcolor{white}\multirow{-2}{*}{\textbf{Dataset}} &
        \cellcolor{white}\multirow{-2}{*}{\textbf{Metric}} &
        \cellcolor{white}\multirow{-2}{*}{\mbox{\textbf{\textcolor{oursAccent}{\sys}}}} &
        \textbf{NCCL} & \textbf{Disk} & \textbf{NCCL} & \textbf{Disk} \\

        \midrule

        \rowcolor{gsmLight}
        \cellcolor{gsmDark!14} & E2E time $\downarrow$ & \cellcolor{oursRow}\textbf{2.82\thinspace s} & 11.28\thinspace s & 51.35\thinspace s & 13.00\thinspace s & 25.07\thinspace s \\
        \rowcolor{gsmLight}
        \cellcolor{gsmDark!14} & GPU $\downarrow$ & \cellcolor{oursRow}\textbf{56.05\thinspace GB} & 61.91\thinspace GB & 60.98\thinspace GB & 67.97\thinspace GB & 68.46\thinspace GB \\
        \rowcolor{gsmLight}
        \cellcolor{gsmDark!14}\multirow{-3}{*}{\textbf{\textcolor{gsmDark}{GSM8K}}} & CPU $\downarrow$ & \cellcolor{oursRow}\textbf{3.45\thinspace GB} & 34.16\thinspace GB & 22.21\thinspace GB & 36.53\thinspace GB & 36.96\thinspace GB \\

        \addlinespace[3pt]

        \rowcolor{humanLight}
        \cellcolor{humanDark!14} & E2E time $\downarrow$ & \cellcolor{oursRow}\textbf{2.86\thinspace s} & 11.26\thinspace s & 50.14\thinspace s & 15.94\thinspace s & 27.21\thinspace s \\
        \rowcolor{humanLight}
        \cellcolor{humanDark!14} & GPU $\downarrow$ & \cellcolor{oursRow}\textbf{57.92\thinspace GB} & 60.67\thinspace GB & 60.75\thinspace GB & 66.95\thinspace GB & 68.02\thinspace GB \\
        \rowcolor{humanLight}
        \cellcolor{humanDark!14}\multirow{-3}{*}{\textbf{\textcolor{humanDark}{HumanEval}}} & CPU $\downarrow$ & \cellcolor{oursRow}\textbf{3.45\thinspace GB} & 34.08\thinspace GB & 24.72\thinspace GB & 36.53\thinspace GB & 36.96\thinspace GB \\

        \addlinespace[3pt]

        \rowcolor{mathLight}
        \cellcolor{mathDark!14} & E2E time $\downarrow$ & \cellcolor{oursRow}\textbf{2.77\thinspace s} & 11.08\thinspace s & 48.22\thinspace s & 20.47\thinspace s & 25.01\thinspace s \\
        \rowcolor{mathLight}
        \cellcolor{mathDark!14} & GPU $\downarrow$ & \cellcolor{oursRow}\textbf{57.82\thinspace GB} & 61.89\thinspace GB & 61.11\thinspace GB & 66.86\thinspace GB & 68.15\thinspace GB \\
        \rowcolor{mathLight}
        \cellcolor{mathDark!14}\multirow{-3}{*}{\textbf{\textcolor{mathDark}{MATH-500}}} & CPU $\downarrow$ & \cellcolor{oursRow}\textbf{3.48\thinspace GB} & 34.16\thinspace GB & 25.27\thinspace GB & 36.09\thinspace GB & 36.63\thinspace GB \\

        \addlinespace[3pt]

        \rowcolor{logiqaLight}
        \cellcolor{logiqaDark!14} & E2E time $\downarrow$ & \cellcolor{oursRow}\textbf{2.87\thinspace s} & 11.28\thinspace s & 48.70\thinspace s & 15.44\thinspace s & 26.51\thinspace s \\
        \rowcolor{logiqaLight}
        \cellcolor{logiqaDark!14} & GPU $\downarrow$ & \cellcolor{oursRow}\textbf{56.06\thinspace GB} & 62.02\thinspace GB & 61.14\thinspace GB & 67.28\thinspace GB & 68.51\thinspace GB \\
        \rowcolor{logiqaLight}
        \cellcolor{logiqaDark!14}\multirow{-3}{*}{\textbf{\textcolor{logiqaDark}{LogiQA}}} & CPU $\downarrow$ & \cellcolor{oursRow}\textbf{3.47\thinspace GB} & 34.13\thinspace GB & 25.27\thinspace GB & 36.43\thinspace GB & 36.84\thinspace GB \\

        \bottomrule
    \end{tabular}

    \vspace{3pt}

    \parbox{\linewidth}{%
        \footnotesize
        \textit{Notes.}
        E2E denotes end-to-end synchronization latency.
        Memory consumption is reported in decimal GB.
        Lower values are better.
    }
\end{table*}

On Qwen3-30B-A3B, \sys achieves the lowest derived E2E synchronization latency across all four workloads and all three bandwidth settings.
Its maximum speedups are $18.3\times$ over ByteCheckpoint and $3.2\times$ over PULSE.

Figure~\ref{fig:cross_cluster_weight_overhead} further normalizes weight-synchronization latency by the policy-training time excluding synchronization.
Even at $10$ Gbps, ByteCheckpoint incurs an overhead of $5.75$--$7.02\times$ the training time and PULSE incurs $2.14$--$2.37\times$, whereas \sys reduces this overhead to $0.68$--$0.99\times$ across both models and all workloads.

Transferred data volume explains much of this advantage.
Figure~\ref{fig:zstd_payload_comparison}(a) shows that \sys produces Zstd-compressed sparse files of only $2.53$--$2.87$ GB, whereas ByteCheckpoint's compressed dense files require approximately $44.5$ GB.
Although fewer than $2\%$ of weights change, the sparse-to-dense file-size ratio exceeds the changed-weight fraction because every sparse entry stores both an int32 index and a \texttt{BF16} value.
Even with this index overhead, \sys files are $16.3\times$ smaller on average.

At $0.1$ Gbps, the difference between \sys and PULSE is relatively small because network transfer dominates both sparse methods.
As bandwidth increases, the widening gap is consistent with local delta construction and application accounting for a larger fraction of E2E latency; the speedup over PULSE reaches $3.2\times$ at $10$ Gbps.

\sys also has the lowest peak GPU and CPU memory.
On Qwen3-30B-A3B, it reduces peak GPU memory by at least $40\%$ and peak CPU memory by at least $66\%$ relative to the storage-mediated baselines, consistent with eliminating model-scale historical and intermediate states.

\begin{table*}[t]
    \centering

    \caption{Cross-cluster weight-synchronization results on \textbf{Qwen3-8B}. Lower values are better.}
    \label{tab:qwen3_8b_cross_cluster}

    \setlength{\tabcolsep}{2.5pt}
    \renewcommand{\arraystretch}{1.12}
    \scriptsize
    \normalfont

    {\bfseries (a) Derived E2E synchronization latency}
    \par\vspace{1pt}

    \begin{tabular}{
        @{}
        >{\centering\arraybackslash}p{1.75cm}
        *{3}{>{\centering\arraybackslash}p{1.35cm}}
        @{\hspace{6pt}}
        *{3}{>{\centering\arraybackslash}p{1.35cm}}
        @{\hspace{6pt}}
        *{3}{>{\centering\arraybackslash}p{1.35cm}}
        @{}
    }
        \toprule

        \rowcolor{white}
        & \multicolumn{3}{c}{\textbf{\textcolor{oursAccent}{\sys}}}
        & \multicolumn{3}{c}{ByteCheckpoint}
        & \multicolumn{3}{c}{PULSE} \\

        \rowcolor{subheaderGray}
        \cellcolor{white}\multirow{-2}{*}{\textbf{Dataset}}
        & \textbf{0.1 Gbps} & \textbf{1 Gbps} & \textbf{10 Gbps}
        & \textbf{0.1 Gbps} & \textbf{1 Gbps} & \textbf{10 Gbps}
        & \textbf{0.1 Gbps} & \textbf{1 Gbps} & \textbf{10 Gbps} \\

        \midrule

        \rowcolor{gsmLight}
        \cellcolor{gsmDark!14}\textbf{\textcolor{gsmDark}{GSM8K}} & \cellcolor{oursRow}117.40\thinspace s & \cellcolor{oursRow}20.47\thinspace s & \cellcolor{oursRow}10.78\thinspace s & 2339.98\thinspace s & 281.43\thinspace s & 75.60\thinspace s & 142.64\thinspace s & 36.50\thinspace s & 27.22\thinspace s \\

        \rowcolor{humanLight}
        \cellcolor{humanDark!14}\textbf{\textcolor{humanDark}{HumanEval}} & \cellcolor{oursRow}118.66\thinspace s & \cellcolor{oursRow}21.87\thinspace s & \cellcolor{oursRow}12.20\thinspace s & 2341.15\thinspace s & 280.49\thinspace s & 74.41\thinspace s & 127.20\thinspace s & 38.35\thinspace s & 29.28\thinspace s \\

        \rowcolor{mathLight}
        \cellcolor{mathDark!14}\textbf{\textcolor{mathDark}{MATH-500}} & \cellcolor{oursRow}127.59\thinspace s & \cellcolor{oursRow}22.94\thinspace s & \cellcolor{oursRow}12.48\thinspace s & 2336.93\thinspace s & 278.35\thinspace s & 72.50\thinspace s & 138.99\thinspace s & 37.83\thinspace s & 27.02\thinspace s \\

        \rowcolor{logiqaLight}
        \cellcolor{logiqaDark!14}\textbf{\textcolor{logiqaDark}{LogiQA}} & \cellcolor{oursRow}143.42\thinspace s & \cellcolor{oursRow}23.65\thinspace s & \cellcolor{oursRow}11.67\thinspace s & 2340.49\thinspace s & 279.15\thinspace s & 73.00\thinspace s & 158.78\thinspace s & 41.02\thinspace s & 28.80\thinspace s \\

        \bottomrule
    \end{tabular}

    \par\vspace{-1pt}

    {\bfseries (b) Peak memory consumption}
    \par\vspace{1pt}

    \begin{tabular}{
        @{}
        >{\centering\arraybackslash}p{1.75cm}
        *{2}{>{\centering\arraybackslash}p{2.15cm}}
        @{\hspace{6pt}}
        *{2}{>{\centering\arraybackslash}p{2.15cm}}
        @{\hspace{6pt}}
        *{2}{>{\centering\arraybackslash}p{2.15cm}}
        @{}
    }
        \toprule

        \rowcolor{white}
        & \multicolumn{2}{c}{\textbf{\textcolor{oursAccent}{\sys}}}
        & \multicolumn{2}{c}{ByteCheckpoint}
        & \multicolumn{2}{c}{PULSE} \\

        \rowcolor{subheaderGray}
        \cellcolor{white}\multirow{-2}{*}{\textbf{Dataset}}
        & \textbf{GPU} & \textbf{CPU}
        & \textbf{GPU} & \textbf{CPU}
        & \textbf{GPU} & \textbf{CPU} \\

        \midrule

        \rowcolor{gsmLight}
        \cellcolor{gsmDark!14}\textbf{\textcolor{gsmDark}{GSM8K}} & \cellcolor{oursRow}52.44\thinspace GB & \cellcolor{oursRow}18.65\thinspace GB & 60.98\thinspace GB & 22.21\thinspace GB & 68.46\thinspace GB & 36.96\thinspace GB \\

        \rowcolor{humanLight}
        \cellcolor{humanDark!14}\textbf{\textcolor{humanDark}{HumanEval}} & \cellcolor{oursRow}52.44\thinspace GB & \cellcolor{oursRow}20.37\thinspace GB & 60.75\thinspace GB & 24.72\thinspace GB & 68.02\thinspace GB & 36.96\thinspace GB \\

        \rowcolor{mathLight}
        \cellcolor{mathDark!14}\textbf{\textcolor{mathDark}{MATH-500}} & \cellcolor{oursRow}52.44\thinspace GB & \cellcolor{oursRow}18.82\thinspace GB & 61.11\thinspace GB & 25.27\thinspace GB & 68.15\thinspace GB & 36.63\thinspace GB \\

        \rowcolor{logiqaLight}
        \cellcolor{logiqaDark!14}\textbf{\textcolor{logiqaDark}{LogiQA}} & \cellcolor{oursRow}52.45\thinspace GB & \cellcolor{oursRow}20.48\thinspace GB & 61.14\thinspace GB & 25.27\thinspace GB & 68.51\thinspace GB & 36.84\thinspace GB \\

        \bottomrule
    \end{tabular}

    \par\vspace{2pt}

    \parbox{\linewidth}{%
        \scriptsize
        \raggedright
        \textit{Notes.}
        All methods use their disk-mediated transfer paths.
        Cross-cluster latency is derived as
        \(T_{\mathrm{same,disk}} + T_{\mathrm{upload,p50}} + T_{\mathrm{download,p50}}\).
        Memory values are taken from the corresponding same-cluster disk runs and reported in decimal GB.
    }
\end{table*}

\subsection{Generality Across Model Architectures and Scales}
\label{sec:qwen3_8b_results}

\textbf{Answer to RQ3: the benefits of receiver-ready sparse updates persist on the smaller dense Qwen3-8B model.}
Together with the Qwen3-30B-A3B results, Tables~\ref{tab:qwen3_8b_same_cluster} and~\ref{tab:qwen3_8b_cross_cluster} show that \sys is effective across both dense and mixture-of-experts architectures.

Within a cluster, \sys synchronizes Qwen3-8B in $2.77$--$2.87$ seconds, yielding speedups of up to $4.0\times$ over ByteCheckpoint-NCCL and $7.4\times$ over PULSE-NCCL.
This result is consistent with the benefit of constructing receiver-specific sparse updates and applying them directly to destination shards.

Across clusters, \sys outperforms ByteCheckpoint and PULSE by up to $19.9\times$ and $2.5\times$, respectively.
As on Qwen3-30B-A3B, its advantage over PULSE increases with bandwidth, a trend consistent with local construction and application becoming more prominent after network transfer ceases to dominate.

The top row of Figure~\ref{fig:cross_cluster_weight_overhead} further shows the
impact of synchronization latency on the training loop by normalizing it against
the policy-training time excluding synchronization.
At $10$ Gbps, the overhead of \sys is only $0.92$--$0.99\times$ the training
time across the four workloads, whereas ByteCheckpoint and PULSE still incur
$5.75$--$6.56\times$ and $2.14$--$2.37\times$ overhead, respectively.
Thus, \sys reduces weight synchronization from a dominant bottleneck to a cost
comparable to, or lower than, the remaining training step.

Transferred data volume follows the same pattern.
Figure~\ref{fig:zstd_payload_comparison}(b) shows that \sys reduces the average Zstd-compressed Qwen3-8B artifact size by $19.7\times$ relative to ByteCheckpoint.

\sys also achieves the lowest peak GPU and CPU memory in both settings.
In the same-cluster experiments, its peak CPU memory is only $3.45$--$3.48$ GB, compared with $34.08$--$34.16$ GB for ByteCheckpoint-NCCL and $36.09$--$36.53$ GB for PULSE-NCCL.
These reductions are consistent with avoiding persistent historical snapshots and model-scale intermediates during delta construction and application.

\section{Conclusion}
\label{sec:conclusion}

\sys makes sparse policy synchronization sparse throughout the complete data path.
It reconstructs pre-update weights through streaming AdamW inversion, compares reconstructed and current \texttt{BF16} representations after aligned conversion, remaps changed elements into receiver-local coordinates, and delivers receiver-specific updates through direct or storage-mediated transport.
This design avoids a persistent historical snapshot, complete-model checkpoint intermediates, and receiver-irrelevant broadcast.
The evaluation shows that these optimizations reduce both latency and memory across model architectures, workloads, and deployment settings.
\sys provides same-cluster speedups of up to $7.6\times$ over ByteCheckpoint-NCCL and $7.4\times$ over PULSE-NCCL.
The derived cross-cluster analysis yields speedups of up to $19.9\times$ over ByteCheckpoint and $3.2\times$ over PULSE; at $10$ Gbps, \sys limits synchronization overhead to $0.68$--$0.99\times$ the policy-training time.
These results show that update sparsity yields its full systems benefit only when construction, layout translation, transport, and application avoid model-scale work.

\bibliographystyle{unsrt}
\bibliography{references}

\appendix

\section{Derivation of AdamW Inversion}
\label{app:adamw_derivation}

This appendix derives Equation~\ref{eq:adamw_inverse} from the standard decoupled AdamW update~\cite{loshchilov2019decoupled}.
The derivation assumes element-wise arithmetic, fixed moment coefficients, the stated bias correction, multiplicative decoupled weight decay, and the standard second-moment denominator without an AMSGrad maximum.
The equalities below hold in exact arithmetic; \sys evaluates the inverse in finite-precision FP32 as described in Section~\ref{sec:optimizer_inversion}.

At optimizer step $t$, the bias-corrected first- and second-moment estimates are
\begin{equation}
    \widehat{m}_t
    =
    \frac{m_t}{1-\beta_1^t},
    \qquad
    \widehat{v}_t
    =
    \frac{v_t}{1-\beta_2^t}.
    \label{eq:appendix_corrected_moments}
\end{equation}

With decoupled weight decay, AdamW updates the weight as
\begin{equation}
    \theta_t
    =
    (1-\eta_t\lambda)\theta_{t-1}
    -
    \eta_t
    \frac{
        \widehat{m}_t
    }{
        \sqrt{\widehat{v}_t}+\epsilon
    }.
    \label{eq:appendix_adamw_forward}
\end{equation}

Rearranging the adaptive update term gives
\begin{equation}
    (1-\eta_t\lambda)\theta_{t-1}
    =
    \theta_t
    +
    \eta_t
    \frac{
        \widehat{m}_t
    }{
        \sqrt{\widehat{v}_t}+\epsilon
    }.
    \label{eq:appendix_adamw_rearranged}
\end{equation}

Dividing by $1-\eta_t\lambda$ and substituting Equation~\ref{eq:appendix_corrected_moments} yields
\begin{equation}
    \theta_{t-1}
    =
    \frac{
        \theta_t
        +
        \eta_t
        \dfrac{
            m_t/(1-\beta_1^t)
        }{
            \sqrt{v_t/(1-\beta_2^t)}+\epsilon
        }
    }{
        1-\eta_t\lambda
    },
    \label{eq:appendix_adamw_inverse}
\end{equation}
This relation is the algebraic basis for the numerical reconstruction in Equation~\ref{eq:adamw_inverse}.

\section{MCore and Hugging Face Parameter Layouts}
\label{app:mcore_hf_layouts}

This appendix explains why \sys performs change detection after conversion rather than directly in the optimizer-native MCore layout.
A fused QKV linear layer provides a concrete example; the same principle applies whenever training and inference layouts differ because of sharding, fusion, permutation, or naming.

\subsection{Optimizer-Native MCore Layout}

MCore organizes model parameters for distributed training.
For a tensor-parallel (TP) linear layer, each TP rank stores a shard of the logical weight matrix, and several logical projections may be fused into one physical parameter.
During AdamW optimization, every locally owned FP32 master-weight slice is element-wise aligned with its first- and second-moment states.
This alignment makes the optimizer-native MCore layout the natural space for the reconstruction in Section~\ref{sec:optimizer_inversion}.

However, a local MCore index identifies an element of a physical training tensor rather than necessarily one logical inference parameter.
Such an index cannot be applied directly by an inference worker without accounting for fusion, permutation, and sharding.

\subsection{Canonical Hugging Face Layout}

The MCore-to-Hugging-Face converter defines a canonical, inference-compatible parameter space.
Each logical inference weight appears as a separately named tensor with a fixed element order.
For the QKV layer considered below, the canonical tensors are
\begin{equation}
    W_Q \in \mathbb{R}^{(h_q d_h)\times d},
    \qquad
    W_K \in \mathbb{R}^{(h_{kv} d_h)\times d},
    \qquad
    W_V \in \mathbb{R}^{(h_{kv} d_h)\times d},
    \label{eq:canonical_qkv_shapes}
\end{equation}
where $d$ is the input hidden dimension, $d_h$ is the head dimension, $h_q$ is the number of query heads, and $h_{kv}$ is the number of key/value heads.
This representation corresponds to the separate \texttt{q\_proj}, \texttt{k\_proj}, and \texttt{v\_proj} parameters in the Hugging Face model interface.

The canonical space need not match the receiver's physical shard layout.
Instead, it provides a common logical coordinate system in which the previous and current versions have identical names, shapes, and element orders.
The transfer plan in Section~\ref{sec:converter_aligned_remapping} maps changes from this common space into receiver-local coordinates.

\subsection{Example: Fused QKV in Grouped-Query Attention}

Consider grouped-query attention (GQA) with $h_{kv}\mid h_q$, where each key/value head is shared by
\begin{equation}
    g = \frac{h_q}{h_{kv}}
    \label{eq:gqa_group_size}
\end{equation}
query heads. For clarity, let $Q_i$, $K_j$, and $V_j$ denote the $d_h\times d$ row blocks associated with query head $i$, key head $j$, and value head $j$, respectively.

A common MCore fused-QKV organization groups the $g$ query heads associated with each key/value head and packs them into one physical tensor:
\begin{equation}
    W_{\mathrm{QKV}}^{\mathrm{MCore}}
    =
    \operatorname{concat}_{j=0}^{h_{kv}-1}
    \left[
        Q_{jg}, Q_{jg+1}, \ldots, Q_{(j+1)g-1}, K_j, V_j
    \right].
    \label{eq:mcore_fused_qkv}
\end{equation}
The concatenation is along the output-row dimension. Hence,
\begin{equation}
    W_{\mathrm{QKV}}^{\mathrm{MCore}}
    \in
    \mathbb{R}^{((h_q+2h_{kv})d_h)\times d}.
    \label{eq:mcore_fused_qkv_shape}
\end{equation}

Under TP degree $P$, MCore further stores this fused tensor as local shards:
\begin{equation}
    W_{\mathrm{QKV},r}^{\mathrm{MCore}}
    =
    W_{\mathrm{QKV}}^{\mathrm{MCore}}
    [\mathcal{R}_r,:],
    \qquad r\in\{0,\ldots,P-1\},
    \label{eq:mcore_tp_shard}
\end{equation}
where $\mathcal{R}_r$ is the output-row range assigned to TP rank $r$.
With a distributed optimizer, data-parallel ranks may own disjoint master-weight and moment slices for a TP-local tensor.
Each owner reconstructs its aligned slice, after which the optimizer group gathers the slices needed to form
$W_{\mathrm{QKV},r}^{\mathrm{MCore}}$ for conversion.

In contrast, the Hugging-Face-compatible representation separates the fused
tensor into three logical parameters:
\begin{equation}
    W_Q =
    \operatorname{concat}_{i=0}^{h_q-1}[Q_i],
    \qquad
    W_K =
    \operatorname{concat}_{j=0}^{h_{kv}-1}[K_j],
    \qquad
    W_V =
    \operatorname{concat}_{j=0}^{h_{kv}-1}[V_j].
    \label{eq:canonical_qkv_unpacking}
\end{equation}
Therefore, the rows of $W_Q$ are not necessarily contiguous in
$W_{\mathrm{QKV}}^{\mathrm{MCore}}$: query blocks are interleaved with their
corresponding key and value blocks. A flat index in the fused MCore tensor
thus cannot be interpreted directly as a flat index of \texttt{q\_proj},
\texttt{k\_proj}, or \texttt{v\_proj}.

To obtain the canonical tensors, the converter gathers the required TP-local pieces, unpacks the fused QKV groups according to Equation~\ref{eq:mcore_fused_qkv}, and concatenates blocks of the same projection according to Equation~\ref{eq:canonical_qkv_unpacking}.
Abstractly, for either the gathered reconstructed or current MCore tensor
$W_{\tau}^{\mathrm{MCore}}$, where $\tau\in\{t-1,t\}$, the converter produces
\begin{equation}
    \mathcal{C}(W_{\tau}^{\mathrm{MCore}})
    =
    \left\{
        (\texttt{q\_proj}, W_{\tau,Q}),
        (\texttt{k\_proj}, W_{\tau,K}),
        (\texttt{v\_proj}, W_{\tau,V})
    \right\}.
    \label{eq:qkv_converter}
\end{equation}

Applying the same converter to both versions makes their comparison layout aligned.
The resulting \texttt{BF16} mask uses logical parameter names and canonical indices rather than the packed training layout.
Receiver-space remapping then partitions or replicates those changes for the destination layout, allowing each receiver to update its local shards directly.

\end{document}

%% file: macro.tex
\usepackage{natbib}
\usepackage{latexsym}

\usepackage{url}
\usepackage{amssymb}
\usepackage[utf8]{inputenc}
\usepackage{microtype}
\usepackage{booktabs}
\usepackage{pifont} 
\usepackage{multirow}
\usepackage{makecell}
\usepackage{xspace}
\usepackage{color}
\usepackage{xcolor}
\usepackage{colortbl}
\usepackage{adjustbox}
\usepackage{hyperref} 
\usepackage[edges]{forest}
\usepackage{tikz} 
\usepackage{caption}
\usepackage{amsfonts}

\hypersetup{
    colorlinks,
    linkcolor={blue!80!black},
    citecolor={blue!80!black},
}
\tikzset{
    root/.style =             {align=center, text width=1cm, rounded corners=3pt, line width=0.3mm, fill=gray!10, draw=gray!80, font=\small},
    demographic/.style =         {align=center, text width=1.8cm, rounded corners=3pt, line width=0.3mm, fill=blue!10, draw=blue!80, font=\footnotesize},
    demographic_work/.style =    {align=center, text width=10cm, rounded corners=3pt, line width=0.3mm, fill=blue!10, draw=blue!0, font=\footnotesize},
    character/.style =         {align=center, text width=1.8cm, rounded corners=3pt, line width=0.3mm, fill=red!10, draw=red!80, font=\footnotesize},
    character_work/.style =    {align=center, text width=10cm, rounded corners=3pt, line width=0.3mm, fill=red!10, draw=red!0, font=\footnotesize},
    personalization/.style =           {align=center, text width=1.8cm, rounded corners=3pt, line width=0.3mm, fill=cyan!10, draw=cyan!80, font=\footnotesize},
    personalization_work/.style =      {align=center, text width=10cm, rounded corners=3pt, line width=0.3mm, fill=cyan!10, draw=cyan!0, font=\footnotesize},
    risk/.style =         {align=center, text width=1.8cm, rounded corners=3pt, line width=0.3mm, fill=orange!10, draw=orange!80, font=\footnotesize},
    risk_work/.style =    {align=center, text width=10cm, rounded corners=3pt, line width=0.3mm, fill=orange!10, draw=orange!0, font=\footnotesize},
}

\usepackage{CJK}

%% file: references.bib
@misc{cursor2026realtimeRL,
  title        = {Real-Time Reinforcement Learning for Composer},
  author       = {{Cursor}},
  year         = {2026},
  howpublished = {\url{https://cursor.com/blog/real-time-rl-for-composer}},
  note         = {Accessed: 2026-07-01}
}

@inproceedings{hybridflow2025,
  title     = {{HybridFlow}: A Flexible and Efficient {RLHF} Framework},
  author    = {Sheng, Guangming and Zhang, Chi and Ye, Zilingfeng and Wu, Xibin and Zhang, Wang and Zhang, Ru and Peng, Yanghua and Lin, Haibin and Wu, Chuan},
  booktitle = {Proceedings of the Twentieth European Conference on Computer Systems},
  pages     = {1279--1297},
  year      = {2025},
  publisher = {ACM},
  doi       = {10.1145/3689031.3696075}
}

@misc{openrlhf2024,
  title        = {OpenRLHF: An Easy-to-use, Scalable and High-performance RLHF Framework},
  author       = {{OpenRLHF Team}},
  year         = {2024},
  howpublished = {\url{https://github.com/OpenRLHF/OpenRLHF}},
  note         = {Accessed: 2026-07-01}
}

@article{roll2025,
  title   = {Reinforcement Learning Optimization for Large-Scale Learning: An Efficient and User-Friendly Scaling Library},
  author  = {Wang, Weixun and Xiong, Shaopan and Chen, Gengru and Gao, Wei and Guo, Sheng and He, Yancheng and Huang, Ju and Liu, Jiaheng and Li, Zhendong and Li, Xiaoyang and Liu, Zichen and Zhao, Haizhou and An, Dakai and Cao, Lunxi and Cao, Qiyang and Deng, Wanxi and Du, Feilei and Gu, Yiliang and Li, Jiahe and Li, Xiang and Liu, Mingjie and Luo, Yijia and Liu, Zihe and Wang, Yadao and Wang, Pei and Wu, Tianyuan and Wu, Yanan and Zhao, Yuheng and Zhao, Shuaibing and Yang, Jin and Yang, Siran and Tan, Yingshui and Yi, Huimin and Xu, Yuchi and Yuan, Yujin and Zhang, Xingyao and Qu, Lin and Su, Wenbo and Wang, Wei and Wang, Jiamang and Zheng, Bo},
  journal = {arXiv preprint arXiv:2506.06122},
  year    = {2025}
}

@misc{slime2026delta,
  title        = {slime Documentation: Delta Weight Sync},
  author       = {{slime Team}},
  year         = {2026},
  howpublished = {\url{https://github.com/THUDM/slime/blob/main/docs/en/advanced/delta-weight-sync.md}},
  note         = {Accessed: 2026-07-01}
}

@inproceedings{loshchilov2019decoupled,
  title     = {Decoupled Weight Decay Regularization},
  author    = {Loshchilov, Ilya and Hutter, Frank},
  booktitle = {International Conference on Learning Representations},
  year      = {2019},
  url       = {https://openreview.net/forum?id=Bkg6RiCqY7}
}

@article{yan2026next,
  title={Next-Generation Agentic Reinforcement Learning Systems Enable Self-Evolving Agents},
  author={Yan, Ran and Fu, Wei and Li, Jiale and Xu, Shusheng and Mei, Zhiyu and Gao, Jiaxuan and Zhang, Jiarui and Shen, Xujie and Dai, Hao and He, Chuyi and others},
  journal={arXiv preprint arXiv:2607.01120},
  year={2026}
}

@article{zhang2026prorl,
  title = {{ProRL Agent}: Rollout-as-a-Service for {RL} Training of Multi-Turn {LLM} Agents},
  author={Zhang, Hao and Liu, Mingjie and Zhang, Shaokun and Han, Songyang and Hu, Jian and Jin, Zhenghui and Zhang, Yuchi and Diao, Shizhe and Lu, Ximing and Xu, Binfeng and others},
  journal={arXiv preprint arXiv:2603.18815},
  year={2026}
}

@inproceedings{gao2026rollart,
  title     = {{RollArt}: Disaggregated {Multi-Task} Agentic {RL} Training at Scale},
  author    = {Gao, Wei and Zhao, Yuheng and Wu, Tianyuan and Xiong, Shaopan and Wang, Weixun and An, Dakai and Cao, Lunxi and Muhtar, Dilxat and Liu, Zichen and Zhao, Haizhou and Huang, Ju and Yang, Siran and Li, Yongbin and Su, Wenbo and Wang, Jiamang and Qu, Lin and Zheng, Bo and Wang, Wei},
  booktitle = {20th USENIX Symposium on Operating Systems Design and Implementation (OSDI 26)},
  pages     = {863--881},
  year      = {2026},
  publisher = {USENIX Association},
  month     = jul,
  url       = {https://www.usenix.org/conference/osdi26/presentation/gao}
}

@article{jiang2026rollout,
  title={Rollout-Training Co-Design for Efficient LLM-Based Multi-Agent Reinforcement Learning},
  author={Jiang, Zhida and Xing, Zhaolong and Lu, Jiawei and Niu, Yipei and Sang, Qingyuan and Zhang, Liangxu and Dai, Wenquan and Shu, Junhua and Wang, Jiaxing and Pei, Qiangyu and others},
  journal={arXiv preprint arXiv:2602.09578},
  year={2026}
}

@article{miahi2026understanding,
  title={Understanding and Exploiting Weight Update Sparsity for Communication-Efficient Distributed RL},
  author={Miahi, Erfan and Belilovsky, Eugene},
  journal={arXiv preprint arXiv:2602.03839},
  year={2026}
}

@article{ruan2026rl,
  title={RL over Commodity Networks: Overcoming the Bandwidth Barrier with Lossless Sparse Deltas},
  author={Ruan, Chaoyi and Luo, Geng and Wan, Xinyi and Zhao, Long and Wang, Qinghe and Zhu, Jiaan and Xu, Duling and Xu, Guanbin and Wei, Dehui and Liu, Xiang and others},
  journal={arXiv preprint arXiv:2602.11456},
  year={2026}
}

@inproceedings{wan2025bytecheckpoint,
  title     = {{ByteCheckpoint}: A Unified Checkpointing System for Large Foundation Model Development},
  author    = {Wan, Borui and Han, Mingji and Sheng, Yiyao and Peng, Yanghua and Lin, Haibin and Zhang, Mofan and Lai, Zhichao and Yu, Menghan and Zhang, Junda and Song, Zuquan and Liu, Xin and Wu, Chuan},
  booktitle = {22nd USENIX Symposium on Networked Systems Design and Implementation (NSDI 25)},
  pages     = {559--578},
  year      = {2025},
  publisher = {USENIX Association},
  month     = apr,
  url       = {https://www.usenix.org/conference/nsdi25/presentation/wan-borui}
}

@inproceedings{megatronlm,
  title     = {Efficient Large-Scale Language Model Training on {GPU} Clusters Using {Megatron-LM}},
  author    = {Narayanan, Deepak and Shoeybi, Mohammad and Casper, Jared and LeGresley, Patrick and Patwary, Mostofa and Korthikanti, Vijay Anand and Vainbrand, Dmitri and Kashinkunti, Prethvi and Bernauer, Julie and Catanzaro, Bryan and Phanishayee, Amar and Zaharia, Matei},
  booktitle = {Proceedings of the International Conference for High Performance Computing, Networking, Storage and Analysis},
  year      = {2021},
  pages = {1--15},
  doi   = {10.1145/3458817.3476209},
  url       = {https://arxiv.org/abs/2104.04473}
}

@article{fsdp,
  title         = {{PyTorch FSDP}: Experiences on Scaling Fully Sharded Data Parallel},
  author        = {Zhao, Yanli and Gu, Andrew and Varma, Rohan and Luo, Liang and Huang, Chien-Chin and Xu, Min and Wright, Less and Shojanazeri, Hamid and Ott, Myle and Shleifer, Sam and Desmaison, Alban and Balioglu, Can and Damania, Pritam and Nguyen, Bernard and Chauhan, Geeta and Hao, Yuchen and Mathews, Ajit and Li, Shen},
  journal       = {arXiv preprint arXiv:2304.11277},
  year          = {2023},
  eprint        = {2304.11277},
  archivePrefix = {arXiv},
  primaryClass  = {cs.DC},
  url           = {https://arxiv.org/abs/2304.11277}
}

@inproceedings{vllm,
  title     = {Efficient Memory Management for Large Language Model Serving with {PagedAttention}},
  author    = {Kwon, Woosuk and Li, Zhuohan and Zhuang, Siyuan and Sheng, Ying and Zheng, Lianmin and Yu, Cody Hao and Gonzalez, Joseph E. and Zhang, Hao and Stoica, Ion},
  booktitle = {Proceedings of the 29th Symposium on Operating Systems Principles},
  pages     = {611--626},
  year      = {2023},
  doi       = {10.1145/3600006.3613165},
  url       = {https://arxiv.org/abs/2309.06180}
}

@inproceedings{sglang,
  title     = {{SGLang}: Efficient Execution of Structured Language Model Programs},
  author    = {Zheng, Lianmin and Yin, Liangsheng and Xie, Zhiqiang and Sun, Chuyue and Huang, Jeff and Yu, Cody Hao and Cao, Shiyi and Kozyrakis, Christos and Stoica, Ion and Gonzalez, Joseph E. and Barrett, Clark and Sheng, Ying},
  booktitle = {Advances in Neural Information Processing Systems},
  volume    = {37},
  year      = {2024},
  doi = {10.52202/079017-2000},
  url       = {https://arxiv.org/abs/2312.07104}
}

@article{fu2025areal,
  title={Areal: A large-scale asynchronous reinforcement learning system for language reasoning},
  author={Fu, Wei and Gao, Jiaxuan and Shen, Xujie and Zhu, Chen and Mei, Zhiyu and He, Chuyi and Xu, Shusheng and Wei, Guo and Mei, Jun and Wang, Jiashu and others},
  journal={Advances in Neural Information Processing Systems},
  volume={38},
  pages={36256--36282},
  year={2025}
}

@article{nemoaligner2024,
  title         = {{NeMo-Aligner}: Scalable Toolkit for Efficient Model Alignment},
  author        = {Shen, Gerald and Wang, Zhilin and Delalleau, Olivier and Zeng, Jiaqi and Dong, Yi and Egert, Daniel and Sun, Shengyang and Zhang, Jimmy and Jain, Sahil and Taghibakhshi, Ali and Sanz Ausin, Markel and Aithal, Ashwath and Kuchaiev, Oleksii},
  journal       = {arXiv preprint arXiv:2405.01481},
  year          = {2024},
  eprint        = {2405.01481},
  archivePrefix = {arXiv},
  primaryClass  = {cs.CL},
  doi           = {10.48550/arXiv.2405.01481},
  url           = {https://arxiv.org/abs/2405.01481}
}

@article{laminar2025,
  title         = {{Laminar}: A Scalable Asynchronous {RL} Post-Training Framework},
  author        = {Sheng, Guangming and Tong, Yuxuan and Wan, Borui and Zhang, Wang and Jia, Chaobo and Wu, Xibin and Wu, Yuqi and Li, Xiang and Zhang, Chi and Peng, Yanghua and Lin, Haibin and Liu, Xin and Wu, Chuan},
  journal       = {arXiv preprint arXiv:2510.12633},
  year          = {2025},
  eprint        = {2510.12633},
  archivePrefix = {arXiv},
  primaryClass  = {cs.LG},
  doi           = {10.48550/arXiv.2510.12633},
  url           = {https://arxiv.org/abs/2510.12633}
}

@article{dora2026,
  title         = {{DORA}: A Scalable Asynchronous Reinforcement Learning System for Language Model Training},
  author        = {Hu, Tianhao and Liu, Xiangcheng and Miao, Yuchun and Xiao, Youshao and Zang, Hongyu and Zheng, Yang and Huang, Xuan and Ding, Jinrui and Zhang, Yufei and Yang, Yu and Zhang, Yi-Kai and Sun, Yueqing and Han, Chengcheng and Ma, Xiandi and Wang, Wei and Gu, Qi and Sun, Yerui and Xie, Yuchen and Cai, Xunliang},
  journal       = {arXiv preprint arXiv:2604.26256},
  year          = {2026},
  eprint        = {2604.26256},
  archivePrefix = {arXiv},
  primaryClass  = {cs.LG},
  doi           = {10.48550/arXiv.2604.26256},
  url           = {https://arxiv.org/abs/2604.26256}
}

@inproceedings{weave2026,
  title     = {{Weave}: Efficient Co-Scheduling for Disaggregated {RL} Post-Training},
  author    = {Wu, Tianyuan and Cao, Lunxi and Wei, Yining and Gao, Wei and Zhao, Yuheng and An, Dakai and Xiong, Shaopan and Lv, Zhiqiang and Huang, Ju and Yang, Siran and Yu, Yinghao and Wang, Jiamang and Qu, Lin and Wang, Wei},
  booktitle = {20th USENIX Symposium on Operating Systems Design and Implementation (OSDI 26)},
  pages     = {809--827},
  year      = {2026},
  publisher = {USENIX Association},
  month     = jul,
  url       = {https://www.usenix.org/conference/osdi26/presentation/wu-tianyuan}
}

@inproceedings{universalcheckpoint2025,
  title     = {Universal Checkpointing: A Flexible and Efficient Distributed Checkpointing System for Large-Scale {DNN} Training with Reconfigurable Parallelism},
  author    = {Lian, Xinyu and Jacobs, Sam Ade and Kurilenko, Lev and Tanaka, Masahiro and Bekman, Stas and Ruwase, Olatunji and Zhang, Minjia},
  booktitle = {2025 USENIX Annual Technical Conference (USENIX ATC 25)},
  pages     = {1519--1534},
  year      = {2025},
  publisher = {USENIX Association},
  month     = jul,
  url       = {https://www.usenix.org/conference/atc25/presentation/lian}
}

@article{tensorhub2026,
  title         = {{TensorHub}: Scalable and Elastic Weight Transfer for {LLM} {RL} Training},
  author        = {Ye, Chenhao and Zhang, Huaizheng and Han, Mingcong and Zhong, Baoquan and Li, Xiang and Chen, Qixiang and Zhang, Xinyi and Zhang, Weidong and Jiang, Kaihua and Zhang, Wang and Sun, He and Xiao, Wencong and Arpaci-Dusseau, Andrea C. and Arpaci-Dusseau, Remzi H.},
  journal       = {arXiv preprint arXiv:2604.09107},
  year          = {2026},
  eprint        = {2604.09107},
  archivePrefix = {arXiv},
  primaryClass  = {cs.DC},
  doi           = {10.48550/arXiv.2604.09107},
  url           = {https://arxiv.org/abs/2604.09107}
}

@article{rose2026,
  title         = {{ROSE}: Rollout On Serving {GPU}s via Cooperative Elasticity for Agentic {RL}},
  author        = {Gao, Wei and Zhao, Yuheng and Muhtar, Dilxat and An, Dakai and Shang, Xuchun and Wu, Tianyuan and Cao, Lunxi and Xiong, Shaopan and Wang, Weixun and Huang, Ju and Ma, Teng and Yang, Siran and Wang, Jiamang and Qu, Lin and Zheng, Bo and Wang, Wei},
  journal       = {arXiv preprint arXiv:2605.06534},
  year          = {2026},
  eprint        = {2605.06534},
  archivePrefix = {arXiv},
  primaryClass  = {cs.DC},
  doi           = {10.48550/arXiv.2605.06534},
  url           = {https://arxiv.org/abs/2605.06534}
}
